# Robust Low-Complexity Randomized Methods for Locating Outliers in Large Matrices

Xingguo Li, *Student Member, IEEE*, and Jarvis Haupt, *Member, IEEE*

*Abstract*—This paper examines the problem of locating outlier columns in a large, otherwise low-rank matrix, in settings where the data are noisy, or where the overall matrix has missing elements. We propose a randomized two-step inference framework, and establish sufficient conditions on the required sample complexities under which these methods succeed (with high probability) in accurately locating the outliers for each task. Comprehensive numerical experimental results are provided to verify the theoretical bounds and demonstrate the computational efficiency of the proposed algorithm.

## I. INTRODUCTION

In this paper we examine a robust outlier identification problem. Given a data matrix $\mathbf{M} \in \mathbb{R}^{n_1 \times n_2}$B, we assume that $\mathbf{M}$ is approximately low-rank, corrupted by (nominally few) outlier columns. More formally, we suppose that

$$\mathbf{M} \approx \mathbf{L} + \mathbf{C}, \qquad (1)$$

where $\mathbf{L}$ is a rank-$r$ matrix and $\mathbf{C}$ is a column-sparse matrix with $k$ nonzero columns that are interpreted as "outliers" of the subspace spanned by columns of $\mathbf{L}$. Our specific goal is to identify the locations of the nonzero columns of $\mathbf{C}$, *without* necessarily identifying the inliers (or the subspace they span), and our particular interest in this work is in doing so when our observations of $\mathbf{M}$ may be contaminated by additive noise, or when only a subset of elements of $\mathbf{M}$ are available, and $n_1, n_2$ are possibly very large relative to the rank $r$ and the number of outliers $k$.

Our investigation is motivated by a wide class of "big data" applications where the outliers themselves are of interest, such as when identifying malicious responses in collaborative filtering applications [1] or finding anomalous patterns in network traffic [2]. Another example arises in computer vision tasks where the aim is to estimation visually salient regions of images [3]–[5]; recent efforts have shown that column outlier models can be viable for describing salient image regions at the "patch" level [6], making the outlier identification approach germane to saliency map estimation tasks.

Within the context of these so-called robust principal component analysis (PCA) tasks, a number of contemporary methods have been developed, which exploit low-dimensional models within the context of convex inference methods. For example, [7], [8] examine robust PCA problems based on entry-wise sparse corruptions, while [9]–[14] propose methods applicable when outliers are present as entire columns. Despite their provable analytical successes, these methods can be computationally demanding when applied to very large data matrices, and more notably for our purposes here, these existing techniques seek to identify or approximate the low-rank matrix $\mathbf{L}$ or the subspace spanned by its columns. Here, our interest is only in locating the outlier columns, and we seek inference procedures having both low sample and implementation complexities (e.g., to obviate the need to store and process the full data matrix).

### A. Overview of Our Contribution

Our initial investigation along these lines [15], [16] proposed a randomized two-step procedure, called adaptive compressive outlier sensing (ACOS), for locating column outliers of a matrix $\mathbf{M} = \mathbf{L} + \mathbf{C}$ (i.e., in a noise-free setting, where all matrix elements are available). The key innovations associated with this approach were the utilization of dimensionality reduction methods, along the lines of those employed in compressed sensing and related areas [17], [18], and sequential adaptive sensing methods, motivated by [19]–[25], where sampling actions are allowed to depend on previous measurements. In our prior work, we showed that when $k = \mathcal{O}(n_2/r)$ and the low-rank matrix satisfies appropriate incoherence conditions, accurate outlier identification is achievable, with high probability, using a total number of *scalar, linear* measurements of the matrix on the order of $r^2 + k$, times constant and logarithmic factors. Our major contributions here come in the form of extensions of the approach of [15], [16] to settings where the data is corrupted by additive noise, or where the available data are incomplete. In the noisy setting, we describe and analyze a randomized sampling and inference procedure that successfully locates outliers (with high probability) using an effective sampling rate of $\frac{\#\text{obs}}{n_1 n_2} = \mathcal{O}\left(\frac{(r + \log n_2)(n_2/n_\mathbf{L})\mu_\mathbf{V} r \log r}{n_1 n_2} + \frac{\log n_2}{n_1}\right)$; in missing-data settings, we present a procedure that succeeds whp using an effective sampling rate $\frac{\#\text{obs}}{p n_1 n_2} = \mathcal{O}\left(\frac{r \mu_\mathbf{L} \log^2 n_2}{p n_1}\right)$, where $n_\mathbf{L}$ is the number of nonzero columns of $\mathbf{L}$, $p$ is observation rate in the missing-data setting, and $\mu_\mathbf{V}$ and $\mu_\mathbf{L}$ are incoherence parameters (defined in the sequel).

### B. Algorithm for Noisy Observations

We first consider model (1) for noisy observations, i.e.,

$$\mathbf{M} = \mathbf{L} + \mathbf{C} + \mathbf{N}, \qquad (2)$$

Submitted December 7, 2016. The authors are with the Department of Electrical and Computer Engineering at the University of Minnesota – Twin Cities. Author emails: {`lixx1661, jdhaupt`}@umn.edu. The authors graciously acknowledge support from NSF Award CCF-1217751 and DARPA Young Faculty Award N66001-14-1-4047.

**Algorithm 1** Robust Adaptive Compressive Outlier Sensing for Noisy Observations (RACOS-N)

**Input:** $\mathbf{M}$, $\gamma \in (0,1)$, $\lambda, \alpha, \varepsilon_1, \varepsilon_2 > 0$, and $q, m \in [n_1]$
**Initalize:** $\mathbf{\Phi} \in \mathbb{R}^{m \times n_1}$, $\mathbf{\Psi} \in \mathbb{R}^{q \times m}$ and $\mathbf{S} = \mathbf{I}_{:,\mathcal{S}}$, where
  $\mathcal{S} = \{j \in [n_2] : S_j \stackrel{iid}{\sim} \text{Bernoulli}(\gamma) = 1\}$ and $p = |\mathcal{S}|$
**Step 1**
  Collect Measurements: $\mathbf{Y}_{(1)} = \mathbf{\Phi M S}$
  Solve OP: $\{\widehat{\mathbf{L}}, \widehat{\mathbf{C}}\} = \arg\min_{\mathbf{L}, \mathbf{C}} \|\mathbf{L}\|_* + \lambda \|\mathbf{C}\|_{1,2}$
            s.t. $\|\mathbf{Y}_{(1)} - \mathbf{L} - \mathbf{C}\|_F \leq \varepsilon_1$
  Estimate: $\widehat{\mathbf{L}}_{(1)}$ by singular value thresholding operation
            on $\widehat{\mathbf{L}}$, i.e. $\widehat{\mathbf{L}}_{(1)} = \widehat{\mathbf{U}} \mathcal{D}_\alpha(\widehat{\mathbf{\Sigma}}) \widehat{\mathbf{V}}^*$
**Step 2**
  Let: $\widehat{\mathcal{L}}_{(1)}$ be the linear subspace spanned by col's of $\widehat{\mathbf{L}}_{(1)}$
  Set: $\mathbf{P}_{\widehat{\mathcal{L}}_{(1)}^\perp} \triangleq \mathbf{I} - \mathbf{P}_{\widehat{\mathcal{L}}_{(1)}}$
  Collect Measurements: $\mathbf{Y}_{(2)} = \mathbf{\Psi} \, \mathbf{P}_{\widehat{\mathcal{L}}_{(1)}^\perp} (\mathbf{\Phi M})$
  Set: $\widehat{z}_i = \|(\mathbf{Y}_{(2)})_{:,i}\|_2$, if $\|(\mathbf{Y}_{(2)})_{:,i}\|_2 > \varepsilon_2$
**Output:** $\widehat{\mathcal{I}}_\mathbf{C} = \{i : \widehat{z}_i \neq 0\}$

---

**Algorithm 2** Robust Adaptive Compressive Outlier Sensing for Incomplete Observations (RACOS-I)

**Input:** $\mathbf{M}$, $\mathbf{\Omega}$, $\gamma_1, \gamma_2 \in (0,1)$, $\rho$ and $\lambda > 0$
**Initalize:** $\mathbf{\Phi} = \mathbf{I}_{\mathcal{S}_1,:}$ and $\mathbf{S} = \mathbf{I}_{:,\mathcal{S}_2}$, where
  $\mathcal{S}_1 = \{i \in [n_1] : S_i \stackrel{iid}{\sim} \text{Bernoulli}(\gamma_1) = 1\}$, $m = |\mathcal{S}_1|$,
  $\mathcal{S}_2 = \{j \in [n_2] : S_j \stackrel{iid}{\sim} \text{Bernoulli}(\gamma_2) = 1\}$, $\check{n}_2 = |\mathcal{S}_2|$.
**Step 1**
  Collect Measurements: $\mathbf{Y}_{(1)} = \mathbf{\Phi M S}$
  Trimming (Optional):
    **for** $j = 1$ **to** $\check{n}_2$
      **if** # of observed entries of $(\mathbf{Y}_{(1)})_{:,j} > \rho m$
        Select: $\rho m$ entires of $(\mathbf{Y}_{(1)})_{:,j}$ uniformly randomly
        Set : The rest entires of $(\mathbf{Y}_{(1)})_{:,j}$ unobserved
    **end for**
  Set: $\mathbf{\Omega}_{(1)}$ be the set of observed entries of $\mathbf{Y}_{(1)}$
  Solve MP: $\{\widehat{\mathbf{L}}_{(1)}, \widehat{\mathbf{C}}_{(1)}\} = \arg\min_{\mathbf{L}, \mathbf{C}} \|\mathbf{L}\|_* + \lambda \|\mathbf{C}\|_{1,2}$
            s.t. $\mathbf{Y}_{(1)} = \mathbf{P}_{\mathbf{\Omega}_{(1)}}(\mathbf{L} + \mathbf{C})$
**Step 2**
  **for** $j = 1$ **to** $n_2$ **do**
    Let: $\widehat{\mathcal{L}}_{\mathcal{I}_j}$ be subspace spanned by col's of $(\widehat{\mathbf{L}}_{(1)})_{\mathcal{I}_j,:}$
    Set: $\mathbf{P}_{\widehat{\mathcal{L}}_{\mathcal{I}_j}^\perp} \triangleq \mathbf{I} - \mathbf{P}_{\widehat{\mathcal{L}}_{\mathcal{I}_j}}$
    Form: $\widehat{z}_j = \|\mathbf{P}_{\widehat{\mathcal{L}}_{(1)}^\perp}(\mathbf{\Phi M})_{\mathcal{I}_j, j}\|_2$
  **end for**
**Output:** $\widehat{\mathcal{I}}_\mathbf{C} = \{i : \widehat{z}_i \neq 0\}$

---

where $\mathbf{N}$ is a matrix of additive noise. The key insight in our two-step approach here follows our initial work of [15], and can be described qualitatively as follows.

We consider throughout a column-wise compressed version $\mathbf{\Phi M}$ having many fewer rows than $\mathbf{M}$, but which still takes the form of a column-wise corrupted low-rank matrix (with the corrupted columns in the same locations as those of the original matrix). In the first step, we apply an existing robust PCA approach designed to be robust to column outliers – called Outlier Pursuit (OP) [9] – to a matrix comprised of a small random subset of columns of $\mathbf{\Phi M}$. This results, in part, in an estimate of the low-rank component $\mathbf{\Phi L}$ of $\mathbf{\Phi M}$, and we identify the subspace spanned from this estimate by the singular vectors of $\mathbf{\Phi L}$ corresponding to singular values above a specified threshold (this serves to mitigate the effects of the noise in the subspace estimate).

Then, a second step incorporates into the sampling operation a composition of an orthogonal projection onto the orthogonal complement of the learned subspace from the first step and an additional column-wise dimensionality reduction operation. This is designed to remove the low-rank component from $\mathbf{\Phi M}$, and to further reduce the dimension of the acquired data. Finally, outlier identification is performed by identifying the columns of the resulting matrix having sufficiently large residual energies. This approach, called Robust Adaptive Compressive Outlier Sensing for noisy observations (RACOS-N), is summarized as Algorithm 1.

### C. Algorithm for Incomplete Observations

We also consider variants of the outlier identification problem when the matrix $\mathbf{M}$ has missing elements, where

$$\mathbf{M} = \mathbf{P}_\mathbf{\Omega}(\mathbf{L} + \mathbf{C}), \quad (3)$$

and $\mathbf{P}_\mathbf{\Omega}$ is an operator that masks its arguments that are not in the index set $\mathbf{\Omega} \subseteq \{1, 2, \ldots, n_1\} \times \{1, 2, \ldots, n_2\}$.

Here we employ an analogous approach as in the noisy case. Namely, we operate throughout on a column-wise compressed matrix $\mathbf{\Phi M}$, but consider specifically the case where $\mathbf{\Phi}$ is a row submatrix of the $n_1 \times n_1$ identity matrix. In the first step, we apply an existing robust PCA approach designed to be robust to column outliers and missing data – called Manipulator Pursuit (MP) [14], [26] – to a matrix comprised of a small random subset of columns of $\mathbf{\Phi M}$. This results, in part, in an estimate of the low-rank component $\mathbf{\Phi L}$, which we denote by $\widehat{\mathbf{L}}_{(1)}$. An optional trimming procedure can be applied before MP by throwing away some entires randomly. This provides better performance for adversarial outliers and improved sampling complexities [14] (see also our additional discussion in Section II).

Then, the second step entails a missing data variant of the orthogonal projection discussed above. Namely, for each $j \in \{1, 2, \ldots, n_2\}$, we let $\mathcal{I}_j \subset \{1, 2, \ldots, n_1\}$ denote the indices of the observed elements of the $j$-th column of $\mathbf{\Phi M}$. Then, for each $j$, we project the observed subvector of the $j$-th column of $\mathbf{\Phi M}$ onto the orthogonal complement of the column space of a row submatrix of $\widehat{\mathbf{L}}_{(1)}$, indexed also by the rows in $\mathcal{I}_j$. A column is recognized as an outlier if its energy after this orthogonal projection is nonzero. We call the algorithm RACOS for incomplete observation (RACOS-I), and summarize it in Algorithm 2.

### D. Comparison with Existing Works

Popular contemporary models for outlier identification based on robust subspace estimation with convex optimization include outlier pursuit (OP) [9], robust computation of linear models (REAPER) [13], and sparse subspace clustering with outliers (SSC) [11]. This is by no means a comprehensive list; see also [27] for a survey of more classical methods in robust statistics. Here we focus on comparing these contemporary models in terms of the model assumption, recovery performance, computational efficiency, the tolerance of missing entries, etc.

TABLE I
COMPARISON OF DIFFERENT OUTLIER IDENTIFICATION MODELS IN TERMS OF ASSUMPTIONS ON **L** AND **C**, COMPUTATIONAL FORMULATION, EXISTENCE OF GUARANTEES FOR MODELING WITH MISSING ENTRIES, AND WHETHER THE GUARANTEES ARE PROBABILISTIC IN NATURE

| Model | OP | REAPER | SSC |
|---|---|---|---|
| Assump. on **L** | Incoherence | Permeance | Incoherence |
| Assump. on **C** | $k = \mathcal{O}(1/r)$ | Small Alignment | Isotropy |
| Computation | Convex | Convex (Relax.) | Convex |
| Missing Entry | Yes [14], [33] | Unknown | Unknown |
| Prob. Result | No | Both | Yes |

OP assumes low coherence of inliers and the number of outliers to be small. While, it does not require outliers to be isotropic and also has guarantees for observation with missing entries (Manipulator Pursuit (MP), [14]). REAPER forms a convex relaxation of least orthogonal absolute deviations [28] and defines several summary statistics explicitly, including permeance (large evidence of inliers), total inlier residual (small deviation/noise on inliers), and alignment (small collinearity of outliers), to reveal the effectiveness of the model. The analysis of REAPER include both the deterministic model and a (Haystack) random model. SSC also assumes incoherence of inliers and isotropy of outliers, where the incoherence here is slightly different with the incoherence in OP, as SSC considers union of subspaces. Though SSC has an algorithmic extension for incomplete observations [29], neither REAPER nor SSC has theoretical guarantees for incomplete observations with outliers so far. In addition, SSC provides a probabilistic result as it considers random outliers [11].

A summary of the properties of these models are provided in Table I. Due to the relatively intuitive condition on **L** and our interest in the case that only a small number of outliers exist (no alignment or uniformity of **C** is required), we use OP as our underlying model that can also handle the case of incomplete observations. Subsequent to our initial work, a downsampling-based approach of Robust PCA was proposed in [30] for noiseless and complete observation, where the goal includes recovery of the subspace spanned by the columns of the low-rank component. Alternative approaches to the robust subspace recovery problem, e.g., [11], [13], [31], [32], can also be incorporated into our overall approach instead of OP, and could yield potential improvements (e.g., in terms of the structural assumptions under which recovery is guaranteed). Investigations along these lines are left for future effort.

### E. Notation

Bold-face upper-case letters (**M**, **Φ** etc.) are used to denote matrices, bold-face lower-case letters (**x**, **v**, etc.) to denote vectors, and non-bold letters are used to denote scalar parameters or constants. We employ both 'block' and 'math' type notations (e.g., **L**, $L$), where the latter are used to denote variables in the optimization tasks. Given a positive integer $n$, we denote $[n] \triangleq \{1, 2, \ldots, n\}$.

The $\ell_p$ norm of a vector $\mathbf{x} = [\mathrm{x}_1 \ \mathrm{x}_2 \ \ldots \ \mathrm{x}_n]$ is $\|\mathbf{x}\|_p = \left(\sum_{i=1}^n |\mathrm{x}_i|^p\right)^{1/p}$. For a matrix $\mathbf{X}$, we denote the nuclear norm (sum of singular values) by $\|\mathbf{X}\|_*$, the spectral norm (largest singular value) by $\|\mathbf{X}\|_2$, the $\ell_{1,2}$ norm (sum of column $\ell_2$ norms) by $\|\mathbf{X}\|_{1,2}$, and the $\ell_{\infty,2}$ norm (largest column $\ell_2$ norm) by $\|\mathbf{X}\|_{\infty,2}$.

For a rank $r$ matrix **L**, we denote the compact singular value decomposition (SVD) of **L** as $\mathbf{L} = \mathbf{U}\mathbf{\Sigma}\mathbf{V}^*$, where $\mathbf{U} \in \mathbb{R}^{n_1 \times r}$ and $\mathbf{V} \in \mathbb{R}^{n_2 \times r}$ have orthonormal columns, and $\mathbf{\Sigma} \in \mathbb{R}^{r \times r}$ is diagonal with the $i$-th largest singular value $\sigma_i(\mathbf{L})$ of **L** as the $i$-th diagonal element, i.e. $\mathbf{\Sigma}_{ii} = \sigma_i(\mathbf{L})$. We also denote $\mathbf{P}_\mathcal{L}(\mathbf{X}) = \mathbf{U}\mathbf{U}^T\mathbf{X}$ and $\mathbf{P}_{\mathcal{L}^\perp}(\mathbf{X}) = (\mathbf{I}-\mathbf{U}\mathbf{U}^T)\mathbf{X}$ as projection operations that project a matrix $\mathbf{X}$ onto the column space and the orthogonal complement of column space of **L** respectively. The mask operator $\mathbf{P}_\mathbf{\Omega}(\cdot)$ is defined via $(\mathbf{P}_\mathbf{\Omega}(\mathbf{X}))_{ij} = \mathbf{X}_{ij}$, if $(i,j) \in \mathbf{\Omega}$; or 0 if $(i,j) \notin \mathbf{\Omega}$ given a support set $\mathbf{\Omega} \subseteq [n_1] \times [n_2]$.

`MATLAB`-inspired notation is used to denote submatrices; e.g., $\mathbf{I}_{\mathcal{S},:}$ (or $\mathbf{I}_{:,\mathcal{S}}$) is used to denote the submatrix formed by extracting rows (or columns) of $\mathbf{I}$ indexed by $\mathcal{S}$. Likewise, we use $\mathbf{X}_{:,j}$ to denote $j$-th column of $\mathbf{X}$.

## II. Main Results

Here we provide the theoretical guarantees of RACOS for model (2) and (3). The noisy observation setting with a generic additive noise will be discussed first. Then, we specialize this to a setting where the noise is random. Finally, we provide the result for the incomplete observation setting with and without a 'trimming' step.

### A. Preliminary Assumptions

We first introduce two important properties on which our recovery guarantees are based. It is well-known that the decomposition of a matrix into a low-rank component and a sparse component is a ill-posed problem in general. For example, a matrix with only one non-zero entry is both a low-rank and sparse matrix. The first property is a widely adopted notion of "incoherence" in the literature of robust PCA [7]–[9] to overcome such identifiability issues.

**Definition II.1** (Row and Column Incoherence Properties). *Let $\mathbf{L} \in \mathbb{R}^{n_1 \times n_2}$ be a rank $r$ matrix with at most $n_\mathbf{L} \leq n_2$ nonzero columns. Given the compact SVD $\mathbf{L} = \mathbf{U}\mathbf{\Sigma}\mathbf{V}^*$, **L** is said to satisfy the **row incoherence property** with parameter $\mu_\mathbf{U} \in [1, n_1/r]$ if*

$$\max_{i \in [n_1]} \|\mathbf{U}^*\mathbf{e}_i\|_2^2 \leq \mu_\mathbf{U} \frac{r}{n_1},$$

*where $\{\mathbf{e}_i\}$ are canonical basis vectors for $\mathbb{R}^{n_1}$. Likewise, **L** is said to satisfy the **column incoherence property** with parameter $\mu_\mathbf{V} \in [1, n_\mathbf{L}/r]$ if*

$$\max_{j \in [n_2]} \|\mathbf{V}^*\mathbf{e}_j\|_2^2 \leq \mu_\mathbf{V} \frac{r}{n_\mathbf{L}},$$

*where $\{\mathbf{e}_j\}$ are canonical basis vectors for $\mathbb{R}^{n_2}$.*

The second important property is a criteria for the random measurement matrices to preserve the Euclidean norm of any fixed vector with high probability, which we formalize as follows.

**Definition II.2** (Distributional Johnson-Lindenstrauss (JL) Property). *A (random) matrix $\mathbf{\Phi} \in \mathbb{R}^{m \times n}$ is said to satisfy the **distributional JL property** if for any fixed $\mathbf{v} \in \mathbb{R}^n$ and any $\varepsilon \in (0,1)$,*

$$\Pr\left( \, \left| \, \|\mathbf{\Phi v}\|_2^2 - \|\mathbf{v}\|_2^2 \, \right| \geq \varepsilon \|\mathbf{v}\|_2^2 \, \right) \leq 2e^{-mf(\varepsilon)}, \quad (4)$$

*where $f(\varepsilon) > 0$ is a constant depending only on $\varepsilon$ that is specific to the distribution of $\mathbf{\Phi}$.*

### B. Guarantees for Noisy Observations

*1) Structural Assumptions:* Motivated from our work of the noiseless case in [15], we state the *structural conditions* for noisy observations as following:

(**d1**) $\operatorname{rank}(\mathbf{L}) = r < \min\{n_1, n_2\}$,
(**d2**) $\mathbf{L}$ has $n_\mathbf{L} = n_2 - k$ nonzero columns,
(**d3**) $\mathbf{L}$ satisfies the *column incoherence property* with parameter $\mu_\mathbf{V}$,
(**d4**) the condition number of $\mathbf{L}$ satisfies $\kappa = \frac{\sigma_1(\mathbf{L})}{\sigma_r(\mathbf{L})} < \infty$, and
(**d5**) $\mathbf{C}$ has $|\mathcal{I}_\mathbf{C}| = k$ nonzero columns, where $\mathcal{I}_\mathbf{C} \triangleq \{i \in [n_2] : \|\mathbf{P}_{\mathcal{L}^\perp} \mathbf{C}_{:,i}\|_2 > \tau_1 \|\mathbf{C}_{:,i}\|_2\}$ for some constant $\tau_1 \in (0,1)$.

The conditions (**d1**)-(**d3**) are natural for $\mathbf{L}$, and are similar to those imposed in our prior work [15]. Note that $n_\mathbf{L} + k \leq n_2$ in general, though we restrict our attention here to the case $n_\mathbf{L} + k = n_2$. Without loss of generality (w.l.o.g.), we assume that for the inlier columns of $\mathbf{L}$, the corresponding columns of $\mathbf{C}$ are 0, and for outlier columns of $\mathbf{C}$, the corresponding columns of $\mathbf{L}$ are zero. The condition (**d4**) assumes the well-conditioning of the low-rank matrix $\mathbf{L}$, which is a mild assumption in practice, and (**d5**) is a condition on the outlier columns that can be viewed as a slightly stronger version of our analogous property assumed in [15] for the noise-free case. That the residuals of the outlier columns need to be sufficiently large is somewhat intuitive, as the outlier columns projected onto the complement of the low-rank subspace need to be large enough due to the inexact estimate of noisy low-rank subspace. In our analysis, the quantity $\tau_1$ is proportional to the upper bound of the estimation error (in spectral norm) of the low-rank subspace, which is simply 0 under analogous structural assumptions when $\mathbf{N} = \mathbf{0}_{n_1 \times n_2}$. Therefore (**d5**) represents a natural extension of the analogous condition imposed in [15] for the noise-free case.

We also impose conditions on the noise to facilitate exact outlier detection. Indeed if, for example, $\mathbf{N}$ has very large Euclidean norm in some column, then we may confuse it for a true outlier column. To avoid such undesirable situations, we impose several conditions on the noise, and its relationship to $\mathbf{L}$ and $\mathbf{C}$. For notational simplicity, we define

$$\eta_\mathbf{N} = \max_{j \in [n_2]} \|\mathbf{N}_{:,j}\|_2.$$

Then the structural conditions of $\mathbf{N}$ are as following:
(**n1**) $\sigma_r(\mathbf{L}) > \frac{90\sqrt{2\gamma}}{\tau_1} n_2 \eta_\mathbf{N}$, and
(**n2**) $\min_{i \in \mathcal{I}_\mathbf{C}} \|\mathbf{C}_{:,i}\|_2 > \tau_2 \eta_\mathbf{N}$ for some constant $\tau_2$,

where $\gamma \in (0,1)$ is the column subsampling parameter (an input to Algorithm 1). The condition (**n1**) is akin to an SNR assumption, and ensures that all singular values of $\mathbf{L}$ dominate the Euclidean norm of columns of $\mathbf{N}$. Note that the condition (**n1**) may seem somewhat restrictive, but both our analysis and numerical evaluation indicate that $\sigma_r(\mathbf{L}) = \Omega(\sqrt{\gamma} n_2 \eta_\mathbf{N})$ may be necessary for our approach. The condition (**n2**) is a structural one that any outlier column dominates the per-column noise. It is interesting to notice that the conditions (**n1**) and (**n2**) hold somewhat trivially when $\mathbf{N} = \mathbf{0}_{n_1 \times n_2}$.

*2) Generic Recovery Guarantees:* We first provide a description of the singular value hard value thresholding operation. Specifically, let the SVD of $\widehat{\mathbf{L}}$ be $\widehat{\mathbf{L}} = \widehat{\mathbf{U}} \widehat{\mathbf{\Sigma}} \widehat{\mathbf{V}}^*$ with $\widehat{\mathbf{\Sigma}} = \operatorname{diag}(\{\widehat{\sigma}_i\}_{1 \leq i \leq \min\{m, \check{n}_2\}})$, where $\check{n}_2$ is the number of columns of $\mathbf{S}$. By choosing a constant $\alpha$, we then apply a singular value thresholding operation defined as $\mathcal{D}_\alpha(\widehat{\mathbf{\Sigma}}) = \operatorname{diag}(\{f(\widehat{\sigma}_i, \alpha)\}_{1 \leq i \leq \min\{m, \check{n}_2\}})$, where $f(\cdot, \cdot)$ is

$$f(\widehat{\sigma}_i, \alpha) \triangleq \begin{cases} \widehat{\sigma}_i, & \text{if } \widehat{\sigma}_i > \alpha \\ 0, & \text{if } \widehat{\sigma}_i \leq \alpha \end{cases},$$

and the estimate of the low-rank matrix is $\widehat{\mathbf{L}}_{(1)} = \widehat{\mathbf{U}} \mathcal{D}_\alpha(\widehat{\mathbf{\Sigma}}) \widehat{\mathbf{V}}^*$. In the next theorem, we state our main results for outlier identification for observation under the general additive noise model (2).

**Theorem II.1** (Accurate Recovery via RACOS-N). *Suppose $\mathbf{M} = \mathbf{L} + \mathbf{C} + \mathbf{N}$, where $\mathbf{L}$ and $\mathbf{C}$ satisfy the structural conditions (d1)-(d5) with the number of outliers $k$ upper bounded by $k_u$,*

$$k \leq k_u = \frac{1}{3(1 + 1024 \, r\mu_\mathbf{V})} \, n_2. \quad (5)$$

*Let the measurement matrices $\mathbf{\Phi}$ and $\mathbf{\Psi}$ be drawn from any distribution following (4), and for a fixed $\delta \in (0,1)$, suppose that the column subsampling parameter $\gamma$, and the row and column sampling parameters $m$ and $q$, respectively, satisfy*

$$\gamma \geq \max \left\{ \frac{200 \log(\frac{6}{\delta})}{n_\mathbf{L}}, \frac{600(1 + 1024 r \mu_\mathbf{V}) \log(\frac{6}{\delta})}{n_2}, \frac{10 r \mu_\mathbf{V} \log(\frac{6r}{\delta})}{n_\mathbf{L}} \right\}, \quad (6)$$

$$m \geq \frac{5(r+1) + \log(2n_2) + \log \frac{2}{\delta}}{f(1/4)}, \quad (7)$$

$$q \geq \frac{4 \log \frac{2n_2}{\delta}}{f(1/4)}. \quad (8)$$

*Further suppose that $\mathbf{N}$ satisfies the structural conditions (n1) and (n2), where the constant $\tau_2$ satisfies*

$$\tau_1 \tau_2 > 6(\beta + 1)(\tau_1/4 + 1) + 90\sqrt{6\gamma}\beta \kappa n_2, \quad (9)$$

*with a constant $\beta > \sqrt{3}$, and the regularization parameter $\lambda$ in OP satisfies*

$$\lambda = \frac{3\sqrt{1 + 1024 \mu_\mathbf{V} r}}{14\sqrt{\check{n}_2}}, \quad (10)$$

*where $\check{n}_2$ is the number of columns of $\mathbf{S}$. Then there exist a singular value hard thresholding constant $\alpha$ and an $\varepsilon_2$ satisfying*

$$18 \gamma n_2 \eta_\mathbf{N} < \alpha < 54 \gamma n_2 \eta_\mathbf{N}, \quad (11)$$

$$\max_{j \in \mathcal{I}_\mathbf{L}} \|\mathbf{\Psi} \mathbf{P}_{\widehat{\mathcal{L}}_{(1)}^\perp} (\mathbf{\Phi} \mathbf{M}_{:,j})\|_2 < \varepsilon_2 < \min_{i \in \mathcal{I}_\mathbf{C}} \|\mathbf{\Psi} \mathbf{P}_{\widehat{\mathcal{L}}_{(1)}^\perp} (\mathbf{\Phi} \mathbf{M}_{:,i})\|_2, \quad (12)$$

*such that the following claims hold simultaneously with probability at least $1 - 3\delta$:*

(C1) *RACOS-N correctly identifies the salient columns of $C$ (i.e., $\widehat{\mathcal{I}}_C = \mathcal{I}_C$), and*
(C2) *the total number of measurements collected is no greater than $\left(\left(\frac{3}{2}\right)\gamma m + q\right) n_2$.*

It is interesting to note that the sufficient condition (5) on the number of identifiable outliers is of the same order compared with OP [9] and noiseless ACOS [15], which can be as large as a fixed proportion of $n_2$ when both the rank $r$ and column coherence parameter $\mu_\mathbf{V}$ are small. In terms of the sample complexity, we show that our approach succeeds with high probability with effective sampling rate $\frac{\#\text{obs}}{n_1 n_2} = \mathcal{O}\left(\frac{(r+\log n_2)(n_2/n_\mathbf{L})\mu_\mathbf{V} r \log r}{n_1 n_2} + \frac{\log n_2}{n_1}\right)$. This may potentially be much smaller than 1 when, e.g., $r$ is small relative to the problem dimensions.

We also present the performance guarantees for RACOS-N when we simply take the column-wise Euclidean norms in Step2, i.e. $\mathbf{\Psi} = \mathbf{I}$ is an identity matrix, in the following corollary. The analysis follows directly from that of Theorem II.1, thus we omit it here.

**Corollary II.1.** *Suppose all conditions in Theorem II.1 hold, except that $\mathbf{\Psi}$ is an identity matrix, i.e. $q = m$, and the constant $\tau_2$ satisfies (9) with a constant $\beta > 1$. If $\lambda$ satisfies (10), then for $\alpha$ and $\varepsilon_2$ satisying (11) and (12) respectively, the following claims hold simultaneously with probability at least $1 - 2\delta$:*

(C3) *RACOS-N correctly identifies the salient columns of $C$ (i.e., $\widehat{\mathcal{I}}_C = \mathcal{I}_C$), and*
(C4) *the total number of measurements collected is no greater than $mn_2$.*

We can see that the recoverability of RACOS-N in terms of the noise for $\mathbf{\Psi} = \mathbf{I}$ is stronger than that when $\mathbf{\Psi}$ is a random matrix, where we require a smaller lower bound for $\tau_2$, hence smaller lower bound requirement for $\min_{i \in \mathcal{I}_\mathbf{C}} \|\mathbf{C}_{:,i}\|_2$ when $\mathbf{\Psi} = \mathbf{I}$. This is intuitively reasonable since fewer random projections facilitate less ambiguity of the original data. On the other hand, the overall sample complexity for random $\mathbf{\Psi}$ is $\mathcal{O}(\gamma m n_2 + q n_2)$, which is potentially much smaller than $\mathcal{O}(mn_2)$ for $\mathbf{\Psi} = \mathbf{I}$, when $\gamma$ and $q$ are small. This can be viewed as a trade off between the outlier detection performance and the sample complexity. Further improvement of sample complexity can be achieved using multivariate regression [34] and the grouping idea [35], if the grouping structure exists among the outliers.

*3) Observations with Random Noise:* We now consider the observation setting (2) with a random noise $\mathbf{N}$. Specifically, we assume that $\mathbf{N}$ has i.i.d. zero-mean Gaussian entries, in which case we can specify the singular value hard thresholding constant $\alpha$. The following Theorem quantifies the constant for the observation with a Gaussian noise. The proof is provided in Appendix VI-D. The analysis can be extended to other type of random noises, such as subgaussian entries, in a straightforward manner.

**Theorem II.2.** *Suppose $\mathbf{M} = \mathbf{L} + \mathbf{C} + \mathbf{N}$, where $\mathbf{L}$ and $\mathbf{C}$ satisfy the structural conditions (d1)-(d5) with $k$ satisfying (5). Also suppose for any $\delta \in (0, 1)$, (6) holds, and the measurement matrices $\mathbf{\Phi}$ and $\mathbf{\Psi}$ are drawn from a distribution satisfying (4), (7) and (8). Further suppose $\mathbf{N}$ has i.i.d. $\mathcal{N}(0, \sigma^2)$ entries and satisfies conditions (n1) and (n2) with the constant $\tau_2$ satisfying (9). If the regularization parameter $\lambda$ satisfies (10), then for $\alpha$ satisfying*

$$18 C_1 \gamma \sigma n_2 < \alpha < 54 C_2 \gamma \sigma n_2, \quad (13)$$

*where $C_1 = \left(n_1 - \left(8n_1 \log \frac{2n_2}{\delta}\right)^{1/2}\right)^{1/2}$ and $C_2 = \left(n_1 + \left(8n_1 \log \frac{2n_2}{\delta}\right)^{1/2}\right)^{1/2}$, and $\varepsilon_2$ satisfying (12), claims (C1) and (C2) hold simultaneously with probability at least $1 - 4\delta$.*

For completeness, we also present the performance guarantees for RACOS-N with $\mathbf{\Psi} = \mathbf{I}$ under the Gaussian noise $\mathbf{N}$ in the following corollary without proof.

**Corollary II.2.** *Suppose all conditions in Theorem II.2 hold, except that $\mathbf{\Psi}$ is an identity matrix, i.e. $q = m$, and the constant $\tau_2$ satisfies (9). If $\lambda$ satisfies (10), then for $\alpha$ and $\varepsilon_2$ satisfying (13) and (12) respectively, claims (C3) and (C4) hold simultaneously with probability at least $1 - 3\delta$.*

We investigate the implications of these results experimentally in Section IV.

### C. Guarantees for Incomplete Observations

We now consider the "missing data" setting.

*1) Structural Assumptions:* For notational simplicity, we denote $\mathbf{C}_\mathbf{\Omega} = \mathbf{P}_\mathbf{\Omega}(\mathbf{C})$. We state the *structural conditions* for the incomplete observation setting (3) as following (adapted from [15]):

(g1) $\text{rank}(\mathbf{L}) = r$,
(g2) $\mathbf{L}$ has $n_\mathbf{L}$ nonzero columns,
(g3) $\mathbf{L}$ satisfies the *row and column incoherence properties* with parameters $\mu_\mathbf{U}$ and $\mu_\mathbf{V}$ respectively, and
(g4) $\mathbf{C}$ has $|\mathcal{I}_\mathbf{C}| = k$ nonzero columns, where $\mathcal{I}_\mathbf{C} \triangleq \{j \in [n_2] : \forall\ \mathcal{I}^* \subset [n_1] \text{ with } |\mathcal{I}^*| \geq \frac{r\mu_\mathbf{U}\log(2r)}{p}, \|(\mathbf{P}_{\mathcal{L}_{\mathcal{I}^*}^\perp}(\mathbf{C}_\mathbf{\Omega}))_{\mathcal{I}^*, j}\|_2 > 0\}$.

Note that the low-rank matrix $\mathbf{L}$ need to satisfy both column and row incoherence properties due to simultaneous column and row sampling procedure. The condition (g4) is from the fact that we only need to consider those observed entries in outlier columns. For missing entries in outlier columns, we will never be able to recover them exactly.

*2) Generic Recovery Guarantees:* In the following theorem, we state our main result for model (3) using RACOS-I without trimming.

**Theorem II.3** (Accurate Recovery via RACOS-I without Trimming). *Suppose $\mathbf{M} = \mathbf{P}_\mathbf{\Omega}(\mathbf{L}+\mathbf{C})$, where the components $\mathbf{L}$ and $\mathbf{C}$ satisfy the structural conditions (g1)-(g4). Let $\mu_\mathbf{L} = \max(\mu_\mathbf{U}, \mu_\mathbf{V})$. Assume $p$ and $k$ satisfy*

$$p \geq p_l = \frac{C_p \mu_\mathbf{L}^2 r^2 \log^3(4n_\mathbf{L})}{n_1}, \quad (14)$$

$$k \leq k_u = \frac{p^2 n_2 / 3}{p^2 + C_k(1 + \frac{3\sqrt{6}\mu_\mathbf{L} r}{p\sqrt{n_1}})\mu_\mathbf{L}^3 r^3 \log^6(4n_\mathbf{L})}, \quad (15)$$

*for some positive constants $C_p$ and $C_k$. Given any $\delta \in (0, 1/2)$, if $n_\mathbf{L} \geq \frac{\delta e^{8p}}{4}$, the row sampling parameter $\gamma_1$ and column sampling parameter $\gamma_2$ satisfy*

$$\gamma_1 \geq \max\left\{\frac{2r\mu_\mathbf{U}\log(2r)}{n_1 p}, \frac{8\log\frac{4n_\mathbf{L}}{\delta}}{n_1 p}, \frac{10r\mu_\mathbf{U}\log\frac{4r}{\delta}}{n_1}, \frac{162 p_l}{p}\right\}, \quad (16)$$

$$\gamma_2 \geq \max\left\{\frac{200\log(\frac{9}{\delta})}{n_\mathbf{L}}, \frac{10r\mu_\mathbf{V}\log(\frac{9r}{\delta})}{n_\mathbf{L}}, \frac{C_{\gamma_2}(\frac{1}{\delta})^{\frac{1}{5}}}{n_2}, \frac{200\log(\frac{9}{\delta})}{k_u}\right\}, \quad (17)$$

*for some positive constant $C_{\gamma_2}$, and the regularization parameter in MP satisfies*

$$\lambda = \frac{1}{48}\sqrt{\frac{p}{9kr\mu_\mathbf{L}\log^2(4\gamma_2 n_\mathbf{L})}}, \quad (18)$$

*then the following claims hold simultaneously with probability at least $1 - 2\delta$:*

- **(C5)** *RACOS-I correctly identifies the salient columns of $\mathbf{C}$ (i.e., $\widehat{\mathcal{I}}_\mathbf{C} = \mathcal{I}_\mathbf{C}$), and*
- **(C6)** *the total number of measurements collected is no greater than $\frac{3}{2}p\gamma_1 n_1 n_2$.*

We see that the sampling complexity reduces from $pn_1n_2$ for the full model to $O(p\gamma_1 n_1 n_2)$ for RACOS-I, which is significant if $r \ll \max\{n_1, n_2\}$. In terms of computational complexity, RACOS-I reduces the size of the matrix operated in MP from $n_1 n_2$ to $\gamma_1 \gamma_2 n_1 n_2$, and the computational cost in each iteration of MP (in the proximal first order algorithm), dominated by SVD, reduces from $\mathcal{O}(n_1 n_2 \min\{n_1, n_2\})$ to $\mathcal{O}(\gamma_1\gamma_2 n_1 n_2 \min\{\gamma_1 n_1, \gamma_2 n_2\})$. This improvement is significant if $r$ is small. Note that the last terms in both (16) and (17) are the dominating terms, which can be improved by the trimming procedure. In the next theorem, we provide the main result for of RACOS-I with trimming.

**Theorem II.4** (Accurate Recovery via RACOS-I with Trimming)**.** *Let $\varphi = \frac{\rho}{p}$. Suppose $\mathbf{M} = \mathbf{P}_\Omega(\mathbf{L} + \mathbf{C})$, where the components $\mathbf{L}$ and $\mathbf{C}$ satisfy the structural conditions (g1)-(g4). Let $\mu_\mathbf{L} = \max(\mu_\mathbf{U}, \mu_\mathbf{V})$. Assume $p$ and $k$ satisfy*

$$p \geq p_l = C_p\left(1 + \frac{1}{\varphi}\right)\frac{\mu_\mathbf{L}r\log^2(2n_2)}{n_1}, \quad (19)$$

$$k \leq k_u = C_k\frac{\varphi}{1+\varphi\sqrt{\varphi}}\frac{pn_\mathbf{L}}{\mu_\mathbf{L}^{3/2}r^{3/2}\log^3(2n_2)}, \quad (20)$$

*for some positive constants $C_p$ and $C_k$. Given any $\delta \in (0, 1/2)$, if the row sampling parameter $\gamma_1$ and the column sampling parameter $\gamma_2$ satisfy (16) and (17) respectively, and the regularization parameter satisfies*

$$\lambda = \frac{1}{48}\sqrt{\frac{1}{\sqrt{(1+\varphi)r\mu_\mathbf{L}}k\log(n_1 + n_\mathbf{L})}}, \quad (21)$$

*then claims (C5) and (C6) hold simultaneously with probability at least $1 - 2\delta$.*

From Theorem II.3, RACOS-I without trimming reduces the dimension of the matrix operated in MP from $n_1 n_2$ to $\mathcal{O}(\mu_\mathbf{L}^5 r^5 \log^9 n_2/p^3)$. On the other hand, from Theorem II.4, RACOS-I with trimming reduces the dimension of the matrix operated in MP from $n_1 n_2$ to $\mathcal{O}(\mu_\mathbf{L}^{5/2} r^{5/2} \log^5 n_2/p^2)$. It is also demonstrated in [14] that (19) and (20) are close to information-theoretic (minimax) optimal, where the trimming is step is crucial in the analysis. We refer interested reader to [14] for further discussion. Though, RACOS-I with trimming has stronger theoretical guarantees, it has one more parameter $\rho$ to choose. Thus, in practice, we take the trimming as an option to trade off the ease of the algorithmic procedure and performance of sampling complexity. As with the noisy case, we evaluate the implications of these results experimentally in Section IV.

## III. PROOF OF MAIN RESULTS

In this section, we provide the sketch of the proof for Theorem II.1, which is formalized by the following intermediate lemmata. The proofs of the lemmata are deferred to the appendix. The proofs for Theorem II.3 and Theorem II.4 are analogous to the proof for Theorem II.1, and are deferred to the supplemental material.

For notional convenience, we first introduce:

$$\widetilde{\mathbf{M}} \triangleq \mathbf{\Phi}\mathbf{M} = \mathbf{\Phi}\mathbf{L} + \mathbf{\Phi}\mathbf{C} + \mathbf{\Phi}\mathbf{N} = \widetilde{\mathbf{L}} + \widetilde{\mathbf{C}} + \widetilde{\mathbf{N}}, \quad (22)$$

We begin by validating that if conditions (**d1**)-(**d5**) and (**n1**)-(**n2**) hold, then analogous structural conditions also hold for $\widetilde{\mathbf{M}}$ provided that $m$ is sufficiently large. This is stated as Lemma III.1, and we provide the proof in Appendix VI-A.

**Lemma III.1.** *Suppose $\mathbf{M} = \mathbf{L} + \mathbf{C} + \mathbf{N}$, where $\mathbf{L}$ and $\mathbf{C}$ satisfy the structural conditions (d1)-(d5), and $\mathbf{N}$ satisfies conditions (n1) and (n2). Given $\delta \in (0, 1)$, further suppose $\mathbf{\Phi}$ is an $m \times n_1$ matrix drawn from a distribution satisfying the distributional JL property (4) with $m$ satisfying (7), and let $\widetilde{\mathbf{M}} = \widetilde{\mathbf{L}} + \widetilde{\mathbf{C}} + \widetilde{\mathbf{N}}$ be as defined in (22). Then, with probability at least $1 - \delta$, the components $\widetilde{\mathbf{L}}$ and $\widetilde{\mathbf{C}}$ satisfy*

- **($\tilde{d}$1)** $\mathrm{rank}(\widetilde{\mathbf{L}}) = r$,
- **($\tilde{d}$2)** $\widetilde{\mathbf{L}}$ *has $n_\mathbf{L}$ nonzero columns,*
- **($\tilde{d}$3)** $\widetilde{\mathbf{L}}$ *satisfies the column incoherence property with parameter $\mu_\mathbf{V}$,*
- **($\tilde{d}$4)** *condition number of $\widetilde{\mathbf{L}}$ satisfies $\frac{\sigma_1(\widetilde{\mathbf{L}})}{\sigma_r(\widetilde{\mathbf{L}})} \leq \sqrt{3}\kappa$, and*
- **($\tilde{d}$5)** $\widetilde{\mathbf{C}}$ *has $|\mathcal{I}_{\widetilde{\mathbf{C}}}| = k$ nonzero columns, where $\mathcal{I}_{\widetilde{\mathbf{C}}} \triangleq \{i \in [n_2] : \|\mathbf{P}_{\widetilde{\mathcal{L}}_\perp}\widetilde{\mathbf{C}}_{:,i}\|_2 > \tau_1\|\widetilde{\mathbf{C}}_{:,i}\|_2/2\}$ and $\mathcal{I}_{\widetilde{\mathbf{C}}} = \mathcal{I}_\mathbf{C}$.*

*Simultaneously, let $\eta_{\widetilde{\mathbf{N}}} = \max_{j \in [n_2]} \|\widetilde{\mathbf{N}}_{:,j}\|_2$, which satisfies*

$$\frac{4}{5}\eta_\mathbf{N} \leq \eta_{\widetilde{\mathbf{N}}} \leq \frac{6}{5}\eta_\mathbf{N}, \quad (23)$$

*then we further have*

- **($\tilde{n}$1)** $\sigma_r(\widetilde{\mathbf{L}}) > \frac{72\sqrt{2\gamma}}{\tau_1}n_2\eta_\mathbf{N}$, *and*
- **($\tilde{n}$2)** $\min_{i \in \mathcal{I}_\mathbf{C}} \|\widetilde{\mathbf{C}}_{:,i}\|_2 > \frac{4}{5}\tau_2\eta_\mathbf{N}$.

Now suppose conditions ($\tilde{d}$1)-($\tilde{d}$5), ($\tilde{n}$1) and ($\tilde{n}$2) hold. We then establish that the number of columns generated by the column downsampling matrix $\mathbf{S}$ is close to $\gamma n_2$, and Step 1 of Algorithm 1 approximately preserves the column space $\widetilde{\mathcal{L}}$ of $\widetilde{\mathbf{L}}$ such that the contaminated outlier columns have larger residuals than that of the inlier columns after the orthogonal projection onto the complement of the estimated low-rank subspace. This is formalized in Lemma III.2 and we provide the proof in Appendix VI-B.

**Lemma III.2.** *Let $\widetilde{\mathbf{M}} = \widetilde{\mathbf{L}} + \widetilde{\mathbf{C}} + \widetilde{\mathbf{N}}$ be an $m \times n_2$ matrix, where the components $\widetilde{\mathbf{L}}$ and $\widetilde{\mathbf{C}}$ satisfy the conditions ($\widetilde{d}1$)-($\widetilde{d}5$). For any $\delta \in (0,1)$, suppose the column sampling parameter $\gamma$ satisfies (6), and $\widetilde{\mathbf{N}}$ satisfies ($\widetilde{n}1$) and ($\widetilde{n}2$) with $k$ satisfying (5). Further suppose $\tau_2$ satisfies (9). If $\lambda$ satisfies (10), and the singular value hard thresholding constant $\alpha$ satisfies (11), then the following claims hold simultaneously with probability at least $1 - \delta$:*

(I) $\mathbf{S}$ *has $\check{n}_2 \leq (3/2)\gamma n_2$ columns, and*

(II) *for any $i \in \mathcal{I}_{\mathbf{C}}$ and $j \in \mathcal{I}_{\mathbf{L}}$, we have that*

$$\|\mathbf{P}_{\widehat{\mathcal{L}}_{(1)}^\perp}(\widetilde{\mathbf{C}} + \widetilde{\mathbf{N}})_{:,i}\|_2 > \beta \|\mathbf{P}_{\widehat{\mathcal{L}}_{(1)}^\perp}(\widetilde{\mathbf{L}} + \widetilde{\mathbf{N}})_{:,j}\|_2. \quad (24)$$

The last intermediate result is to show that Step 2 of Algorithm 1 produces the correct set of outlier columns with high probability, provided that (24) holds. This is summarized in Lemma III.3 and its proof is provided in Appendix VI-C.

**Lemma III.3.** *For any $\delta \in (0,1)$, suppose (24) holds, and $\mathbf{\Psi} \in \mathbb{R}^{q \times m}$ is a matrix drawn from a distribution satisfying the distributional JL property (4) with $q$ satisfying (8). If $\varepsilon_2$ satisfies (12), then with probability at least $1 - \delta$, we have $\widehat{\mathcal{I}}_{\mathbf{C}} = \mathcal{I}_{\widetilde{\mathbf{C}}}$ from Step 2 of Algorithm 1.*

The overall results of Theorem II.1 follows by combining three intermediate results provided in Lemma III.1, Lemma III.2, and Lemma III.3, using the union bound. Therefore, with probability at least $1 - 3\delta$, the claims (**C1**) and (**C2**) of Theorem II.1 hold.

## IV. EXPERIMENTAL EVALUATION

In this section, we demonstrate explicitly via the numerical evaluation that the sample complexities we derived in the main results are tight in practice[1]. We also examine the computational performance to quantify the improvement of our proposed method over the full data models OP and MP. The timing is recorded as the CPU execution time of the algorithm for different combinations of parameters $(m, \gamma, \gamma_1, \gamma_2)$. All results are evaluated by averaging 100 random trials with MATLAB R2014b on an iMac with a 3.4 GHz Intel Core i7 processor, 32 GB memory, and running OS X 10.8.5.

### A. Evaluation for Noisy Observation Settings

A trial is deemed a success if the following holds:

$$\min_{i \in \mathcal{I}_{\mathbf{C}}} \|\mathbf{\Psi} \, \mathbf{P}_{\widehat{\mathcal{L}}_{(1)}^\perp}(\mathbf{\Phi}\mathbf{M}_{:,i})\|_2 > \max_{i \in \mathcal{I}_{\mathbf{L}}} \|\mathbf{\Psi} \, \mathbf{P}_{\widehat{\mathcal{L}}_{(1)}^\perp}(\mathbf{\Phi}\mathbf{M}_{:,i})\|_2,$$

which implies that there exists a constant threshold $\varepsilon_2$ such that the column-wise hard thresholding yields accurate support recovery. For the singular value hard thresholding, we choose the constant $\alpha$ that preserves 99% of the sum of singular values, which performs well in our settings of evaluations.

---
[1]Outlier recovery transition plots for ACOS are provided in Li & Haupt [15] to demonstrate the recoverability in term of $r$ and $k$ for different levels of noise $\sigma_{\mathbf{N}}$ for noisy observations and different sampling parameter $p$ for observation with missing entries. More results on real data evaluations are provided in [35], [36].

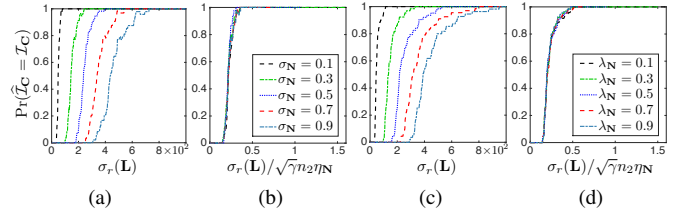

Fig. 1. Demonstration of the probability of success versus the minimal singular value $\sigma_r(\mathbf{L})$ of $\mathbf{L}$ for Gaussian noise under different choices of the variance $\sigma_{\mathbf{N}}$ (a and b) and Laplace noise under different choices of the parameters $\lambda_{\mathbf{N}}$ (c and d). (b) and (d) provide the results with rescaling of $\sigma_r(\mathbf{L})$ by $\sqrt{\gamma}n_2\eta_{\mathbf{N}}$.

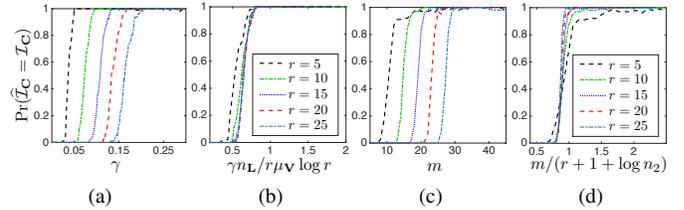

Fig. 2. Demonstration of the probability of success versus column subsample parameter $\gamma$ (a and b) and row sampling parameter $m$ (c and d) for noisy observations under different settings of rank $r$ of $\mathbf{L}$. (b) and (d) provide the results with rescaling of $\gamma$ by $\frac{r\mu_{\mathbf{V}}\log(r)}{n_{\mathbf{L}}}$ and $m$ by $r+1+\log k$ respectively.

We generate both the row sampling matrix $\mathbf{\Phi}$ and the row reduction matrix $\mathbf{\Psi}$ with i.i.d. $\mathcal{N}(0,1)$ entries, without the evaluation of $\mathbf{\Psi} = \mathbf{I}$. We fix $n_1 = 100$, $n_2 = 1000$, $q = 20$, $k = 0.2n_2$, $n_{\mathbf{L}} = n_2 - k$, and $\lambda = 0.4$, and justify the claimed bounds via varying parameters, such as $r$, $m$ and $\gamma$.

We first demonstrate that $\sigma_r(\mathbf{L}) = \Omega(\sqrt{\gamma}n_2\eta_{\mathbf{N}})$ in (**n1**) appears to be a necessary bound in practice. Let $r = 5$, $m = 0.3n_1$ and $\gamma = 0.2$. We generate two random matrices $\mathbf{U} \in \mathbb{R}^{n_1 \times r}$ and $\mathbf{V} \in \mathbb{R}^{n_{\mathbf{L}} \times r}$ with i.i.d. $\mathcal{N}(0,1)$ entries, and take $\mathbf{L}_0 = [\mathbf{U}\mathbf{V}^T \; \mathbf{0}_{n_1 \times k}]$. Then let $\mathbf{L} = \frac{\sigma_r(\mathbf{L})}{\sigma_r(\mathbf{L}_0)} \mathbf{U}_0\mathbf{\Sigma}_0\mathbf{V}_0^T$, where $\mathbf{U}_0\mathbf{\Sigma}_0\mathbf{V}_0^T$ is SVD of $\mathbf{L}_0$, $\sigma_r(\mathbf{L}_0) = (\mathbf{\Sigma}_0)_{rr}$ is the minimal singular value of $\mathbf{L}_0$, and $\sigma_r(\mathbf{L})$ is a parameter to control the singular values of $\mathbf{L}$. The outlier matrix is generated as $\mathbf{C} = [\mathbf{0}_{n_1 \times n_{\mathbf{L}}} \; \mathbf{W}]$ where $\mathbf{W} \in \mathbb{R}^{n_1 \times k}$ has i.i.d. $\mathcal{N}(0,r)$ entries. We evaluate two type of noises in this section: (1) the noise matrix $\mathbf{N}$ has i.i.d. $\mathcal{N}(0, \sigma_{\mathbf{N}}^2)$ entries with five different values of $\sigma_{\mathbf{N}} \in \{0.1, 0.3, 0.5, 0.7, 0.9\}$; (2) $\mathbf{N}$ has i.i.d. zero-mean Laplace entries with five difference choices of parameters $\lambda_{\mathbf{N}} \in \{0.1, 0.3, 0.5, 0.7, 0.9\}$. Note that $\mathbf{U}$, $\mathbf{V}$, $\mathbf{W}$ and $\mathbf{N}$ are mutually independent. For each $\sigma_{\mathbf{N}}$ and $\lambda_{\mathbf{N}}$, we choose $\sigma_r(\mathbf{L}) \in \{2, 4, 6, \ldots\}$ and demonstrate the empirical values of $\Pr(\widehat{\mathcal{I}}_{\mathbf{C}} = \mathcal{I}_{\mathbf{C}})$ (over 100 trials) in Figure 1, with and without the rescaling of $\sigma_r(\mathbf{L})$ by $\sqrt{\gamma}n_2\eta_{\mathbf{N}}$.

In panel (a), we observe that as $\sigma_{\mathbf{N}}$ increases, the threshold of $\sigma_r(\mathbf{L})$ for correct identification of outlier columns with high probability also increases, as we expect. On the other hand, when we rescale $\sigma_r(\mathbf{L})$ by $\sqrt{\gamma}n_2\eta_{\mathbf{N}}$ in panel (b), all curves corresponding to different values of $\sigma_{\mathbf{N}}$ are aligned together. Besides, when the ratio $\frac{\sigma_r(\mathbf{L})}{\sqrt{\gamma}n_2\eta_{\mathbf{N}}}$ goes beyond 1, the probability of correct outlier detection is 1, which verifies our assumption (**n1**) in this case. Analogous results are observed for Laplace noise as well.

Next, we evaluate the bound of the column subsampling parameter $\gamma$ w.r.t. the rank $r$ in (6). We fix $\mathbf{N}$ as Gaussian noise with i.i.d. entries with $\sigma_{\mathbf{N}} = 0.01$ and the following discussion. Let $m = 0.3n_1$. We generate $\mathbf{L} = [\mathbf{UV}^T \ \mathbf{0}_{n_1 \times k}]$ and $\mathbf{C} = [\mathbf{0}_{n_1 \times n_{\mathbf{L}}} \ \mathbf{W}]$, where $\mathbf{U} \in \mathbb{R}^{n_1 \times r}$ and $\mathbf{V} \in \mathbb{R}^{n_{\mathbf{L}} \times r}$ have i.i.d. $\mathcal{N}(0,1)$ entries and $\mathbf{W} \in \mathbb{R}^{n_1 \times k}$ has i.i.d. $\mathcal{N}(0, r)$ entries. We choose five values of ranks $r \in \{5, 10, 15, 20, 25\}$, and plot the empirical probability of correct outlier identification $\Pr(\widehat{\mathcal{I}}_{\mathbf{C}} = \mathcal{I}_{\mathbf{C}})$ versus the column subsampling parameter $\gamma \in \{0.001, 0.002, 0.003, \ldots, 0.3\}$ for each $r$ in Figure 2 (a,b). When $r$ increases, the column subsampling parameter $\gamma$ also needs to increase for correct outlier identification with high probability. If we normalize $\gamma$ with $\frac{r\mu_{\mathbf{V}} \log r}{n_{\mathbf{L}}}$, which is generally the dominating term in (6), then all curves corresponding to different ranks $r$ align together, as shown in panel (b). Further, high probability of success is achieved when the ratio $\gamma / \frac{r\mu_{\mathbf{V}} \log r}{n_{\mathbf{L}}} > 1$, as we have established in (6).

Analogous evaluation for the bound of the row sampling parameter $m$ w.r.t. $r$ in (7) is also provided. Let $\gamma = 0.2$, and the generations of $\mathbf{L}$, $\mathbf{C}$ and $\mathbf{N}$ are identical to those in the previous evaluation for $\gamma$. Again, we choose five values of ranks $r \in \{5, 10, 15, 20, 25\}$, and plot $\Pr(\widehat{\mathcal{I}}_{\mathbf{C}} = \mathcal{I}_{\mathbf{C}})$ versus $m \in \{1, 2, 3, \ldots, 50\}$ for each $r$ in Figure 2. The observation matches with the bound (7) that increasing $m$ facilitates the accurate recovery for increasing $r$, and the ratio $m/(r+1+\log n_2) > 1$ facilitates correct recovery with high probability, as shown in panel (c).

### B. Evaluation for Incomplete Observation Settings

We proceed all experiments here using the trimming option. The setting is as follows. A trial is claimed to be a success if $\widehat{\mathcal{I}}_{\mathbf{C}} = \mathcal{I}_{\mathbf{C}}$, where $\widehat{\mathcal{I}}_{\mathbf{C}}$ is given in Algorithm 2 for RACOS-I. We fix $n_1 = 100$, $n_2 = 1000$, $k = 0.2n_2$ and $\lambda = 0.4$. The generations of $\mathbf{L}$ and $\mathbf{C}$ follow that in the previous evaluation for noisy observations. We apply the trimming step by choosing $\rho = 0.9$ throughout.

First, we evaluate the bound (16) for $\gamma_1$ w.r.t. the rank $r$ and the entry-wise sampling parameter $p$ respectively. In evaluating $\gamma_1$ w.r.t. $r$, let $\gamma_2 = 0.2$, $p = 0.5$, and the rank be chosen from $r \in \{3, 6, 9, 12, 15\}$. In evaluating $\gamma_1$ w.r.t. $p$, we fix $\gamma_2 = 0.2$, $r = 5$, and choose the sampling parameter from $p \in \{0.4, 0.5, 0.6, 0.7, 0.9\}$. For each $p$ and $r$, we set $\gamma_1 \in \{0.02, 0.04, 0.06, \ldots, 1\}$ and demonstrate the plots of $\Pr(\widehat{\mathcal{I}}_{\mathbf{C}} = \mathcal{I}_{\mathbf{C}})$ versus $\gamma_1$ in Figure 3 (top row). In panel (a), we observe that when $r$ increases, the threshold of $\gamma_1$ for correct identification with high probability also increases due to the positive dependence of $\gamma_1$ and $r$. Analogously in panel (c), as $p$ increases, the threshold of $\gamma_1$ for correct identification with high probability decreases due to the inverse dependence of $\gamma_1$ on $p$. On the other hand, in panel (b) and (d), when we rescale $\gamma_1$ by $\frac{\mu_{\mathbf{L}} r \log(n_2)}{n_1 p}$, which is the dominating term of (16) in our setting, all curves corresponding to different values of $p$ align gracefully and facilitates high probability of recovery with the ratio $> 1$.

We carry out the similar evaluation of the bound (17) for $\gamma_2$ w.r.t. $r$ and $p$ respectively. We follow the same settings stated above and plot $\Pr(\widehat{\mathcal{I}}_{\mathbf{C}} = \mathcal{I}_{\mathbf{C}})$ versus $\gamma_2$ in Figure 3 (bottom row). The observation is that increasing $\gamma_2$ facilitates the accurate recovery for increasing $r$ and decreasing $p$, and the ratio $\gamma_2 / \frac{\mu_{\mathbf{L}} r \log(n_2)}{n_{\mathbf{L}} p} > 1$ corresponds to correct detection with high probability. However, we do not have explicitly $\gamma_2 = \Omega(\frac{\mu_{\mathbf{L}} r \log(n_2)}{n_{\mathbf{L}} p})$ in our bound (17), where the dominating term, considering (20), is $\gamma_2 = \Omega(\frac{\mu_{\mathbf{L}}^{3/2} r^{3/2} \log^3(n_2)}{n_{\mathbf{L}} p})$. This suggests that further improvement may be achieved in terms of the sampling complexity of $\gamma_2$ in (20), which we leave for future investigation.

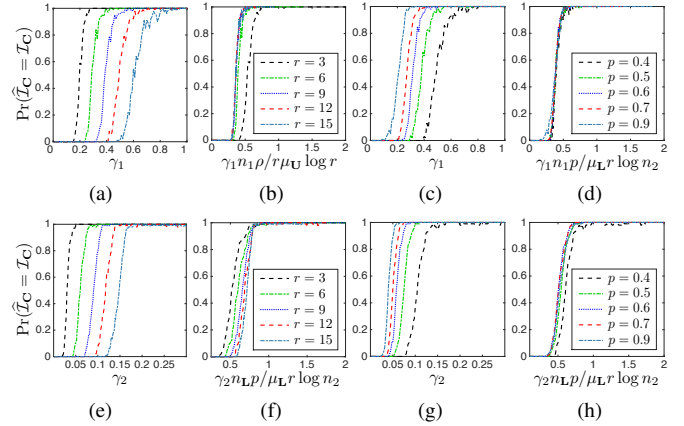

Fig. 3. Demonstration of the probability of success versus the row subsampling parameter $\gamma_1$ (top row) and the column subsampling parameter $\gamma_2$ (bottom row) under different settings of the rank $r$ (a,b,e,f) and the entry-wise sampling parameter $p$ (c,d,g,h). Panel (b) and (d) provide the results with rescaling of $\gamma_1$ by $\frac{\mu_{\mathbf{L}} r \log(n_2)}{n_1 p}$. Panel (f) and (h) provide the results with rescaling of $\gamma_2$ by $\frac{\mu_{\mathbf{L}} r \log(n_2)}{n_{\mathbf{L}} p}$.

### C. Timing Performance

We further examine the timing performances for both models. We fix $n_1 = 500$, $n_2 = 1000$, $k = 0.2n_2$, $n_{\mathbf{L}} = n_2 - k$, and $\lambda = 0.4$, and generate $\mathbf{L}$, $\mathbf{C}$, and the Gaussian noise $\mathbf{N}$ in the same way described above.

For the noisy observation setting, we fix $r = 10$ and choose different combinations of the row sampling parameter $m$ and the column sapling parameter $\gamma$. More specifically, we choose $m \in \{10, 20, 30, \ldots, 500\}$ and $\gamma \in \{0.02, 0.04, 0.06, \ldots, 1\}$, where the pair $(m, \gamma) = (500, 1)$ corresponds to operating on the full-size data matrix $\mathbf{M}$. We first provide the "phase transition" behavior as discussed in [15] for all combinations of $m$ and $\gamma$ and a fixed $\lambda = 0.5$ in OP. Then we record the CPU execution time of Algorithm 1. The phase transition and the contour plot of timing evaluation are provided in Figure 4 (a,b). The values on contour lines are the speedups of algorithm compared with the full size model, i.e. $(m, \gamma) = (500, 1)$. We can see that our approach shows significant advantage in terms of computational efficiency over the full data model when $m$ and $\gamma$ are small. For example, when $(m/n_1, \gamma) = (0.1, 0.1)$, our approach is $> 100$ times faster than that using the full data. Another interesting observation is that the full size model $(m, \gamma) = (500, 1)$ is not the slowest here, while the nearly full size model is the slowest. This is because in the full data model, we do not need to construct the random projection matrices and the corresponding projection operations. In applications, such as the salient image feature detection, speedup of over 100 times can be achieved with comparable performances [15], [35].

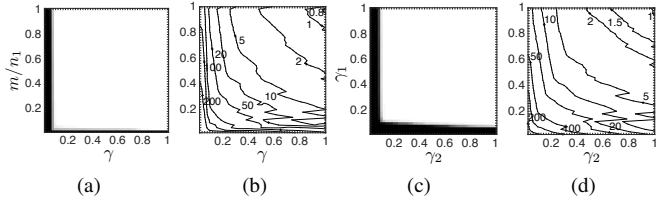

Fig. 4. Demonstration of the performance using different combinations of $m$ and $\gamma$ for noisy observations and different combinations of $\gamma_1$ and $\gamma_2$ for incomplete observations via (a,c) phase transition and (b,d) timing evaluation of OP/MP respectively.

Analogous evaluation is also carried out for the incomplete observation setting. We fix $p = 0.4$ and $r = 5$, and choose different combinations of the row sampling parameter $\gamma_1 \in \{0.02, 0.04, 0.06, \ldots, 1\}$ and the column sampling parameter $\gamma_2 \in \{0.02, 0.04, 0.06, \ldots, 1\}$, where the pair $(\gamma_1, \gamma_2) = (1, 1)$ corresponds to the full-size data model. We provide the phase transition and the contour plot of timing evaluation of Algorithm 2 for each pair of $(\gamma_1, \gamma_2)$ in Figure 4 (c,d). Significant improvement of the computational efficiency over the full data model is also observed. For example, when $(\gamma_1, \gamma_2) = (0.2, 0.2)$, our approach is $> 50$ times faster than the full data model.

## V. DISCUSSION

The idea of identifying outliers from a few linear summaries of the original data matrix may be extended to a large class of models. The key insight is that if the structure of the problem can be (approximately) preserved, then significantly improved computational complexity may be achieved by operating on a much smaller dimensional problem. This is also closely related with the recent development of sketching techniques in linear algebra, data mining, and machine learning [37]. Our future interest includes, but not limited to, extension of a more challenging observation model in the existence of both missing entries and noise, and the identification of outliers from union of subspaces.

## VI. APPENDIX

### A. Proof of Lemma III.1

We start with demonstrating (23), $(\widetilde{n}1)$ and $(\widetilde{n}2)$. We state clearly here that all following results are obtained by taking $\varepsilon = \sqrt{2}/4$. The choice of $\sqrt{2}/4$ is somewhat arbitrary and we choose this fixed value for concreteness. Note that given $\varepsilon \in (0, 1)$ and $\delta \in (0, 1)$, if $\boldsymbol{\Phi}$ satisfies the distributional JL property with $m$ specified in (7), then $\boldsymbol{\Phi}$ is an $\varepsilon$-stable embedding of $\left(\mathcal{L}, \cup_{i \in \mathcal{I}_\mathbf{C}} \{\mathbf{C}_{:,i} + \mathbf{N}_{:,i}\} \cup_{j \in [n_2]} \{\mathbf{N}_{:,j}\} \cup \{\mathbf{0}\}\right)$ with probability at least $1 - \delta$ (Lemma III.1 in [15]). This implies $\sqrt{1-\varepsilon}\|\mathbf{N}_{:,i}\|_2 \leq \|\widetilde{\mathbf{N}}_{:,i}\|_2 \leq \sqrt{1+\varepsilon}\|\mathbf{N}_{:,i}\|_2$ for any $i \in [n_2]$, which results in (23).

We also have,
$$\sigma_r(\widetilde{\mathbf{L}}) \overset{(i)}{\geq} \sqrt{1-\varepsilon}\sigma_r(\mathbf{L}) \overset{(ii)}{>} \sqrt{1-\varepsilon}\frac{90\sqrt{2\gamma}}{\tau_1}n_2\eta_\mathbf{N} \overset{(iii)}{\geq} \frac{72\sqrt{2\gamma}}{\tau_1}n_2\eta_\mathbf{N},$$

where $(i)$ is a direct application of Theorem 1 in [38] via the $\varepsilon$-stable embedding property, $(ii)$ is from $(\mathbf{n}1)$, and $(iii)$ is from (23), which results in $(\widetilde{n}1)$.

To verify $(\widetilde{n}2)$, we have from the $\varepsilon$-stable embedding property of $\boldsymbol{\Phi}$ and $(\mathbf{n}2)$ that for any $i \in \mathcal{I}_\mathbf{C}$, $j \in [n_2]$,
$$\|\widetilde{\mathbf{C}}_{:,i}\|_2 \geq \sqrt{1-\varepsilon}\|\mathbf{C}_{:,i}\|_2 > \sqrt{1-\varepsilon}\tau_2\|\mathbf{N}_{:,j}\|_2.$$

Next, we demonstrate $(\widetilde{d}1)$-$(\widetilde{d}5)$. $(\widetilde{d}1)$-$(\widetilde{d}3)$ follow directly from the result in [15] (Lemma III.1). We have from Theorem 1 in [38],
$$\frac{\sigma_1(\widetilde{\mathbf{L}})}{\sigma_r(\widetilde{\mathbf{L}})} \leq \frac{\sqrt{1+\varepsilon}\sigma_1(\mathbf{L})}{\sqrt{1-\varepsilon}\sigma_r(\mathbf{L})} \leq \sqrt{3}\kappa, \quad (25)$$

which establish $(\widetilde{d}4)$. For $(\widetilde{d}5)$, we have
$$\|\mathbf{P}_{\widetilde{\mathcal{L}}^\perp}\widetilde{\mathbf{C}}_{:,i}\|_2 \overset{(i)}{\geq} \sqrt{\frac{1-2\varepsilon}{1-\varepsilon}}\|\mathbf{P}_{\mathcal{L}^\perp}\mathbf{C}_{:,i}\|_2 \overset{(ii)}{>} \tau_1\sqrt{\frac{1-2\varepsilon}{1-\varepsilon}}\|\mathbf{C}_{:,i}\|_2$$
$$\overset{(iii)}{\geq} \tau_1\sqrt{\frac{1-2\varepsilon}{1-\varepsilon^2}}\|\widetilde{\mathbf{C}}_{:,i}\|_2 \geq \tau_1\|\widetilde{\mathbf{C}}_{:,i}\|_2/2,$$

where $(i)$ and $(iii)$ are from $\boldsymbol{\Phi}$ being an $\varepsilon$-stable embedding, and $(ii)$ is from $(\mathbf{d}5)$. It is straightforward from the definition of the orthogonal projection $\mathcal{P}_{\widetilde{\mathcal{L}}}$ that for any $i \in \mathcal{I}_\mathbf{C}$ and $j \in \mathcal{I}_\mathbf{L}$,
$$\|\mathbf{P}_{\widetilde{\mathcal{L}}^\perp}(\widetilde{\mathbf{L}} + \widetilde{\mathbf{N}})_{:,j}\|_2 = \|\mathbf{P}_{\widetilde{\mathcal{L}}^\perp}\widetilde{\mathbf{N}}_{:,j}\|_2 < \|\widetilde{\mathbf{N}}_{:,j}\|_2$$
$$\overset{(i)}{<} \sqrt{3}\|\widetilde{\mathbf{C}}_{:,i}\|_2/\tau_2 \overset{(ii)}{<} \tau_1\|\widetilde{\mathbf{C}}_{:,i}\|_2/2,$$

where $(i)$ is from (23) and $(\widetilde{n}2)$, and $(ii)$ is from the fact $\tau_2 > 2\sqrt{3}/\tau_1$ from (9). By definition of $\mathcal{I}_{\widetilde{\mathbf{C}}}$, we have $\mathcal{I}_{\widetilde{\mathbf{C}}} = \mathcal{I}_\mathbf{C}$.

### B. Proof of Lemma III.2

For notional convenience, we introduce
$$\check{\mathbf{M}} \triangleq \widetilde{\mathbf{M}}\mathbf{S} = \widetilde{\mathbf{L}}\mathbf{S} + \widetilde{\mathbf{C}}\mathbf{S} + \widetilde{\mathbf{N}}\mathbf{S} = \check{\mathbf{L}} + \check{\mathbf{C}} + \check{\mathbf{N}},$$

where $\mathbf{S}$ is the column sampling matrix. From Lemma III.2 in [15], the following results hold with probability at least $1 - \delta$ when $\mathbf{S}$ is generated as specified[2]:

(**a1**) $\mathbf{S}$ has $(1/2)\gamma n_2 \leq \check{n}_2 \leq (3/2)\gamma n_2$ columns,

(**a2**) $\check{\mathbf{L}}$ has $\check{n}_\mathbf{L} \leq (3/2)\gamma n_\mathbf{L}$ nonzero columns,

(**a3**) $\check{\mathbf{C}}$ has $\check{k} \leq (3/2)\gamma k_u$ nonzero columns,

(**a4**) $\sigma_1^2(\widetilde{\mathbf{V}}^*\mathbf{S}) \leq (3/2)\gamma$, and

(**a5**) $\sigma_r^2(\widetilde{\mathbf{V}}^*\mathbf{S}) \geq (1/2)\gamma$,

where $\widetilde{\mathbf{V}}$ is the matrix of right singular vectors from the compact SVD of $\widetilde{\mathbf{L}}$, i.e. $\widetilde{\mathbf{L}} = \widetilde{\mathbf{U}}\widetilde{\boldsymbol{\Sigma}}\widetilde{\mathbf{V}}^*$, and $\sigma_i(\widetilde{\mathbf{V}}^*\mathbf{S})$ denotes the $i$-th largest singular value of $\widetilde{\mathbf{V}}^*\mathbf{S}$. Note that parameters $(1/2)$ and $(3/2)$ arising in the conditions (**a1**)-(**a5**) are somewhat arbitrary, and they are fixed to the values here for ease of exposition.

Claim (I) follows directly from (**a1**). To justify Claim (ii), we have (with a minor modification of Lemma III.2 in [15]) that when (**a1**)-(**a5**) and (6) are satisfied, the following structural conditions of $\check{\mathbf{L}}$ and $\check{\mathbf{C}}$ hold:

(**ď1**) $\check{r} = \operatorname{rank}(\check{\mathbf{L}}) = r$,

(**ď2**) $\check{\mathbf{L}}$ has $n_{\check{\mathbf{L}}} \leq \frac{3}{2}\gamma n_\mathbf{L}$ nonzero columns,

(**ď3**) $\check{\mathbf{L}}$ satisfies the *column incoherence property* with parameter $\mu_{\check{\mathbf{V}}} = 9\mu_\mathbf{V}$, and

(**ď4**) $\mathcal{I}_{\check{\mathbf{C}}} \triangleq \{i : \|\mathbf{P}_{\check{\mathcal{L}}^\perp}(\check{\mathbf{C}}_{:,i})\|_2 > 0\}$ with $|\mathcal{I}_{\check{\mathbf{C}}}| = \check{k}$, where $\check{\mathcal{L}}$ denotes the linear subspace spanned by columns of $\check{\mathbf{L}}$

---

[2]Here we use $\delta/6$ instead of $\delta/5$ for bounding the probability of the complement of each event (**a1**)-(**a5**) in the proof of Lemma III.2 of [15], and a different (**a3**) with that in [15].

and $\mathbf{P}_{\check{\mathcal{L}}^\perp}$ is the orthogonal projection operator onto the orthogonal complement of $\check{\mathcal{L}}$ in $\mathbb{R}^m$, and

$$\check{k} \leq \left(\frac{1}{1+(1024/9)\ \check{r}\mu_{\check{\mathbf{V}}}}\right)\check{n}_2. \tag{26}$$

Proofs of (**ď1**)-(**ď3**) are identical to those in Lemma A.4 in [15]. The condition (**ď4**) follows from (**a3**) since

$$\check{k} \leq \frac{3}{2}\gamma k_u \stackrel{(i)}{\leq} \frac{3k_u\check{n}_2}{n_2} \stackrel{(ii)}{=} \left(\frac{1}{1+(1024/9)\ \check{r}\mu_{\check{\mathbf{V}}}}\right)\check{n}_2,$$

where $(i)$ is from (**a1**), and $(ii)$ is from (**ď1**) and (**ď3**).

Now we verify (**a3**), which will be discussed in two cases.

**Case 1**. Since $\gamma \geq \frac{600(1+1024r\mu_{\mathbf{L}})\log(\frac{6}{\delta})}{n_2}$ in (6), we have $\frac{200}{\gamma}\log(\frac{6}{\delta}) \leq \frac{n_2}{3(1+1024r\mu_{\mathbf{V}})}$. Let $k$ satisfy

$$\frac{200}{\gamma}\log(\frac{6}{\delta}) \leq k \leq k_u = \frac{n_2}{3(1+1024r\mu_{\mathbf{V}})}. \tag{27}$$

Note that $\check{k}$ is a Hypergeometric random variable with distributions Hyp$(n_2, \check{n}_2, k)$, which is parameterized by the population size $n_2$, the total number of draws $\check{n}_2$, and the total positive elements $k$. Then we have from [15] (second part of the proof of Lemma III.2) that

$$\Pr\left(\check{k} > (3/2)\gamma k\right) \leq \exp\left(-\gamma k/200\right). \tag{28}$$

Let the R.H.S. of (28) be no larger than $\delta/6$. Then we have $\check{k} \leq (3/2)\gamma k \leq (3/2)\gamma k_u$, i.e., (**a3**) holds, w.p. $\geq 1 - 6/\delta$, if $k$ satisfies (27).

**Case 2**. If $k < \frac{200}{\gamma}\log(\frac{6}{\delta})$, then (**a3**) holds w.p. $\geq 1-6/\delta$ by the stochastic ordering argument [39]. More specifically, let $\check{k}_1$ and $\check{k}_2$ be Hypergeometric random variables with distributions Hyp$(n_2, \check{n}_2, k_1)$ and Hyp$(n_2, \check{n}_2, k_2)$ respectively, where $k_1 > k_2$. Then by Lemma 4.1 in [16], for any $x \in [0, \infty)$,

$$\Pr(\check{k}_2 \leq x) \geq \Pr(\check{k}_1 \leq x).$$

This indicates that when $k < \frac{200}{\gamma}\log(\frac{6}{\delta})$, we have $\check{k} \leq (3/2)\gamma k_u$ w.p. $\geq 1 - 6/\delta$.

Next, we show that the estimate $\widehat{\mathbf{L}}$ of $\check{\mathbf{L}}$ can be obtained with the existence of noise, and $\|\check{\mathbf{L}} - \widehat{\mathbf{L}}\|_2$ can be bounded in terms of $\|\check{\mathbf{N}}\|_{1,2}$. We formalize this notion in Lemma VI.1, and provide the proof in Appendix VI-G.

**Lemma VI.1** (Outlier Pursuit with Noise, adapted from Theorem 2 of [9]). *Let $\check{\mathbf{M}} = \check{\mathbf{L}} + \check{\mathbf{C}} + \check{\mathbf{N}}$ be an $m \times \check{n}_2$ matrix whose components $\check{\mathbf{L}}$ and $\check{\mathbf{C}}$ satisfy the structural conditions (**ď1**)-(**ď4**) with $\check{k}$ satisfying (26). Then for $\lambda = \frac{\sqrt{9+1024\mu_{\check{\mathbf{L}}}r}}{14\sqrt{\check{n}_2}}$ and any solution pair obtained from the* outlier pursuit

$$\{\widehat{\mathbf{L}}, \widehat{\mathbf{C}}\} = \operatorname*{argmin}_{\mathbf{L},\mathbf{C}} \|\mathbf{L}\|_* + \lambda\|\mathbf{C}\|_{1,2}$$
$$s.t.\ \ \|\check{\mathbf{M}} - (\mathbf{L}+\mathbf{C})\|_F \leq \varepsilon_1, \tag{29}$$

*there exists $\check{\mathbf{L}}_0$ and $\check{\mathbf{C}}_0$ such that $\check{\mathbf{M}}_0 = \check{\mathbf{L}}_0 + \check{\mathbf{C}}_0$, where $\check{\mathbf{L}}_0$ has the correct column space of $\check{\mathbf{L}}$, $\check{\mathbf{C}}_0$ has the correct column support of $\check{\mathbf{C}}$, and*

$$\|\check{\mathbf{L}}_0 - \widehat{\mathbf{L}}\|_2 \leq 10\|\check{\mathbf{N}}\|_{1,2},\quad \|\check{\mathbf{C}}_0 - \widehat{\mathbf{C}}\|_2 \leq 9\|\check{\mathbf{N}}\|_{1,2}. \tag{30}$$

Note that we can only guarantee the estimation errors in terms of some $\check{\mathbf{L}}_0$ having the same column space with $\check{\mathbf{L}}$ and some $\check{\mathbf{C}}_0$ sharing the same column support with $\check{\mathbf{C}}$, which is enough for our purpose of analysis. Lemma VI.1 is further utilized to bound $\|\mathbf{P}_{\widehat{\mathcal{L}}_{(1)}} - \mathbf{P}_{\widetilde{\mathcal{L}}}\|_2$ away from 1, where $\widehat{\mathcal{L}}_{(1)}$ is the column space of $\widehat{\mathbf{L}}_{(1)}$, obtained by the singular values thresholding operation of $\widehat{\mathbf{L}}$. We will show that $\widehat{\mathbf{L}}_{(1)}$ has the same rank with $\check{\mathbf{L}}$ from our choices of parameters.

Remind that $\check{\mathbf{L}}$ has zero columns when the corresponding columns of $\check{\mathbf{C}}$ are nonzero. Since $\check{\mathbf{C}}$ and $\check{\mathbf{C}}_0$ have the same column support, thus $\check{\mathbf{L}}_0$ and $\check{\mathbf{L}}$ are identical for non-zero columns of $\check{\mathbf{L}}$. Besides, columns of $\check{\mathbf{L}}_0$ may be non-zero at the locations of zero columns of $\check{\mathbf{L}}$. Therefore, we have that for any $i \in [r]$, $\sigma_i(\check{\mathbf{L}}_0) \geq \sigma_i(\check{\mathbf{L}})$. This can be seen using the following argument. Since $\check{\mathbf{L}}_0$ and $\check{\mathbf{L}}$ have the same column space, then the $i$-th singular values of $\check{\mathbf{L}}_0$ and $\check{\mathbf{L}}$ satisfy

$$\sigma_i(\check{\mathbf{L}}_0) = \|\mathbf{u}_i^T\check{\mathbf{L}}_0\|_2 \geq \|\mathbf{u}_i^T\check{\mathbf{L}}\|_2 = \sigma_i(\check{\mathbf{L}}), \tag{31}$$

where $\mathbf{u}_i$ is the $i$-th left singular value of $\check{\mathbf{L}}_0$. Combining ($\widetilde{\mathbf{n}}$**1**), (**a5**), and (31), we have

$$\sigma_r(\check{\mathbf{L}}_0) \geq \sigma_r(\check{\mathbf{L}}) \geq \sqrt{\frac{\gamma}{2}}\sigma_r(\widetilde{\mathbf{L}}) \geq \frac{72}{\tau_1}\gamma n_2\eta_{\mathbf{N}}. \tag{32}$$

On the other hand, we have

$$\sigma_r(\check{\mathbf{L}}_0) - \sigma_r(\widehat{\mathbf{L}}) \stackrel{(i)}{\leq} \|\check{\mathbf{L}}_0 - \widehat{\mathbf{L}}\|_2 \stackrel{(ii)}{\leq} 10\|\check{\mathbf{N}}\|_{1,2} \stackrel{(iii)}{\leq} 12\check{n}_2\eta_{\mathbf{N}}$$
$$\stackrel{(iv)}{\leq} \frac{12 \cdot 3\gamma n_2 \eta_{\mathbf{N}}}{2} = 18\gamma n_2\eta_{\mathbf{N}}, \tag{33}$$

where $(i)$ is from Lemma III.1 in the supplemental material, $(ii)$ is from (30), $(iii)$ is from (23), and $(iv)$ is from (**a1**). Combining (32), (33), and $\tau_1 \in (0,1)$, we have

$$\sigma_r(\widehat{\mathbf{L}}) \geq \left(\frac{4}{\tau_1} - 1\right)18\gamma n_2\eta_{\mathbf{N}} \geq 54\gamma n_2\eta_{\mathbf{N}} > 0. \tag{34}$$

Using the same argument as (33) and (34), we have

$$\sigma_{r+1}(\widehat{\mathbf{L}}) \leq \sigma_{r+1}(\check{\mathbf{L}}_0) + \|\check{\mathbf{L}}_0 - \widehat{\mathbf{L}}\|_2 \leq 18\gamma n_2\eta_{\mathbf{N}}. \tag{35}$$

When the singular value thresholding constant $\alpha$ satisfies (11), it is guaranteed by (34) and (35) that rank$(\widehat{\mathbf{L}}_{(1)}) = r$, where $\widehat{\mathbf{L}}_{(1)} = \widehat{\mathbf{U}}\mathcal{D}_\alpha(\widehat{\mathbf{\Sigma}})\widehat{\mathbf{V}}^*$ and $\mathcal{D}_\alpha(\widehat{\mathbf{\Sigma}})$ is the hard thresholding operation. Combining (32) and (33), we have

$$\|\mathbf{P}_{\widehat{\mathcal{L}}_{(1)}} - \mathbf{P}_{\widetilde{\mathcal{L}}}\|_2 \stackrel{(i)}{=} \|\mathbf{P}_{\widehat{\mathcal{L}}_{(1)}} - \mathbf{P}_{\check{\mathcal{L}}}\|_2 \stackrel{(ii)}{\leq} \frac{\|\check{\mathbf{L}} - \widehat{\mathbf{L}}_{(1)}\|_2}{\sigma_r(\check{\mathbf{L}})}$$
$$\stackrel{(iii)}{\leq} \frac{\|\check{\mathbf{L}} - \widehat{\mathbf{L}}\|_2}{\sigma_r(\check{\mathbf{L}})} \stackrel{(iv)}{\leq} \frac{18\gamma n_2\eta_{\mathbf{N}}}{\sigma_r(\check{\mathbf{L}})}, \tag{36}$$

where $(i)$ is from $\mathbf{P}_{\widetilde{\mathcal{L}}} = \mathbf{P}_{\check{\mathcal{L}}}$, $(ii)$ is from the additive perturbation bound of the orthogonal projection in Lemma III.2 in the supplemental material, $(iii)$ is from the way we generate $\widehat{\mathbf{L}}_{(1)}$, and $(iv)$ is from (33). Let $\tau_3$ be the R.H.S. of (36). Combining (32) and (36), we have

$$\|\mathbf{P}_{\widehat{\mathcal{L}}_{(1)}} - \mathbf{P}_{\widetilde{\mathcal{L}}}\|_2 \leq \tau_3 \leq \frac{18\gamma n_2\eta_{\mathbf{N}}}{72\gamma n_2\eta_{\mathbf{N}}/\tau_1} = \frac{\tau_1}{4}. \tag{37}$$

Using triangle inequality and Cauchy-Schwarz inequality, we have for any $j \in \mathcal{I}_{\mathbf{L}}$

$$\left|\|\mathbf{P}_{\widetilde{\mathcal{L}}^\perp}(\widetilde{\mathbf{L}}+\widetilde{\mathbf{N}})_{:,j}\|_2 - \|\mathbf{P}_{\widehat{\mathcal{L}}_{(1)}^\perp}(\widetilde{\mathbf{L}}+\widetilde{\mathbf{N}})_{:,j}\|_2\right|$$
$$\leq \|(\mathbf{P}_{\widetilde{\mathcal{L}}^\perp} - \mathbf{P}_{\widehat{\mathcal{L}}_{(1)}^\perp})(\widetilde{\mathbf{L}}+\widetilde{\mathbf{N}})_{:,j}\|_2$$
$$\leq \|\mathbf{P}_{\widetilde{\mathcal{L}}^\perp} - \mathbf{P}_{\widehat{\mathcal{L}}_{(1)}^\perp}\|_2 \cdot \|(\widetilde{\mathbf{L}}+\widetilde{\mathbf{N}})_{:,j}\|_2$$
$$= \|\mathbf{P}_{\widehat{\mathcal{L}}_{(1)}} - \mathbf{P}_{\widetilde{\mathcal{L}}}\|_2 \cdot \|(\widetilde{\mathbf{L}}+\widetilde{\mathbf{N}})_{:,j}\|_2 \leq \tau_3\|(\widetilde{\mathbf{L}}+\widetilde{\mathbf{N}})_{:,j}\|_2,$$

from which we have for any $j \in \mathcal{I}_{\mathbf{L}}$ and $i \in \mathcal{I}_{\mathbf{C}}$,

$$\|\mathbf{P}_{\widehat{\mathcal{L}}_{(1)}^\perp}(\widetilde{\mathbf{L}}+\widetilde{\mathbf{N}})_{:,j}\|_2 \leq \|\mathbf{P}_{\widetilde{\mathcal{L}}^\perp}(\widetilde{\mathbf{L}}+\widetilde{\mathbf{N}})_{:,j}\|_2 + \tau_3\|(\widetilde{\mathbf{L}}+\widetilde{\mathbf{N}})_{:,j}\|_2$$
$$= \|\mathbf{P}_{\widetilde{\mathcal{L}}^\perp}\widetilde{\mathbf{N}}_{:,j}\|_2 + \tau_3\|(\widetilde{\mathbf{L}}+\widetilde{\mathbf{N}})_{:,j}\|_2. \quad (38)$$

Applying the same analysis, we have for any $i \in \mathcal{I}_{\mathbf{C}}$

$$\|\mathbf{P}_{\widehat{\mathcal{L}}_{(1)}^\perp}(\widetilde{\mathbf{C}}+\widetilde{\mathbf{N}})_{:,j}\|_2$$
$$\geq \|\mathbf{P}_{\widetilde{\mathcal{L}}^\perp}(\widetilde{\mathbf{C}}+\widetilde{\mathbf{N}})_{:,i}\|_2 - \tau_3\|(\widetilde{\mathbf{C}}+\widetilde{\mathbf{N}})_{:,i}\|_2. \quad (39)$$

Then for any $j \in \mathcal{I}_{\mathbf{L}}$ and $i \in \mathcal{I}_{\mathbf{C}}$, we have

$$\|\mathbf{P}_{\widetilde{\mathcal{L}}^\perp}(\widetilde{\mathbf{C}}_{:,i})\|_2 \stackrel{(i)}{>} \frac{\tau_1}{2}\|\widetilde{\mathbf{C}}_{:,i}\|_2 \stackrel{(ii)}{>} \frac{\tau_1}{4}\left(\|\widetilde{\mathbf{C}}_{:,i}\|_2 + \frac{4}{5}\tau_2\eta_{\mathbf{N}}\right)$$
$$\stackrel{(iii)}{>} \frac{\tau_1}{4}\|\widetilde{\mathbf{C}}_{:,i}\|_2 + \left(\frac{6}{5}(\beta+1)(\frac{\tau_1}{4}+1) + 18\sqrt{6\gamma}\beta\kappa n_2\right)\eta_{\mathbf{N}}$$
$$\stackrel{(iv)}{\geq} \frac{\tau_1}{4}\|\widetilde{\mathbf{C}}_{:,i}\|_2 + (\beta+1)(\frac{\tau_1}{4}+1)\eta_{\widetilde{\mathbf{N}}} + 18\sqrt{6\gamma}\beta\kappa n_2\eta_{\mathbf{N}}$$
$$\stackrel{(v)}{\geq} \frac{\tau_1}{4}\|\widetilde{\mathbf{C}}_{:,i}\|_2 + (\frac{\tau_1}{4}+1)\|\widetilde{\mathbf{N}}_{:,i}\|_2 + \beta(\frac{\tau_1}{4}+1)\|\widetilde{\mathbf{N}}_{:,j}\|_2$$
$$\quad + 18\sqrt{6\gamma}\beta\kappa n_2\eta_{\mathbf{N}}$$
$$\stackrel{(vi)}{\geq} \frac{\tau_1}{4}\|\widetilde{\mathbf{C}}_{:,i}\|_2 + \frac{\tau_1}{4}\|\widetilde{\mathbf{N}}_{:,i}\|_2 + \beta\|\mathbf{P}_{\widetilde{\mathcal{L}}^\perp}\widetilde{\mathbf{N}}_{:,j}\|_2 + \frac{\beta\tau_1}{4}\|\widetilde{\mathbf{N}}_{:,j}\|_2$$
$$\quad + 18\sqrt{6\gamma}\beta\kappa n_2\eta_{\mathbf{N}} + \|\widetilde{\mathbf{N}}_{:,i}\|_2, \quad (40)$$

where $(i)$ is from ($\widetilde{d}$5), $(ii)$ is from ($\widetilde{n}$2), $(iii)$ is from (9), $(iv)$ is from (23), $(v)$ is from the definition of $\eta_{\widetilde{\mathbf{N}}}$, and $(vi)$ is from the condition $\beta > 1$. Then, we have

$$\|\mathbf{P}_{\widetilde{\mathcal{L}}^\perp}(\widetilde{\mathbf{C}}_{:,i}+\widetilde{\mathbf{N}}_{:,i})\|_2 \geq \|\mathbf{P}_{\widetilde{\mathcal{L}}^\perp}(\widetilde{\mathbf{C}}_{:,i})\|_2 - \|\mathbf{P}_{\widetilde{\mathcal{L}}^\perp}(\widetilde{\mathbf{N}}_{:,i})\|_2$$
$$\stackrel{(i)}{>} \frac{\tau_1}{4}\|\widetilde{\mathbf{C}}_{:,i}\|_2 + \frac{\tau_1}{4}\|\widetilde{\mathbf{N}}_{:,i}\|_2 + \beta\|\mathbf{P}_{\widetilde{\mathcal{L}}^\perp}\widetilde{\mathbf{N}}_{:,j}\|_2 + \frac{\beta\tau_1}{4}\|\widetilde{\mathbf{N}}_{:,j}\|_2$$
$$\quad + 18\sqrt{6\gamma}\beta\kappa n_2\eta_{\mathbf{N}}$$
$$\stackrel{(ii)}{\geq} \frac{\tau_1}{4}\|(\widetilde{\mathbf{C}}+\widetilde{\mathbf{N}})_{:,i}\|_2 + \beta\|\mathbf{P}_{\widetilde{\mathcal{L}}^\perp}\widetilde{\mathbf{N}}_{:,j}\|_2$$
$$\quad + 18\sqrt{\gamma}\beta n_2\eta_{\mathbf{N}}\left(\sqrt{6}\kappa + \frac{\sqrt{\gamma}\|\widetilde{\mathbf{N}}_{:,j}\|_2}{\sigma_r(\check{\mathbf{L}})}\right)$$
$$\stackrel{(iii)}{\geq} \frac{\tau_1}{4}\|(\widetilde{\mathbf{C}}+\widetilde{\mathbf{N}})_{:,i}\|_2 + \beta\|\mathbf{P}_{\widetilde{\mathcal{L}}^\perp}\widetilde{\mathbf{N}}_{:,j}\|_2$$
$$\quad + 18\sqrt{\gamma}\beta n_2\eta_{\mathbf{N}}\left(\frac{\sqrt{2}\sigma_1(\widetilde{\mathbf{L}})}{\sigma_r(\check{\mathbf{L}})} + \frac{\sqrt{\gamma}\|\widetilde{\mathbf{N}}_{:,j}\|_2}{\sigma_r(\check{\mathbf{L}})}\right)$$
$$\stackrel{(iv)}{\geq} \frac{\tau_1}{4}\|(\widetilde{\mathbf{C}}+\widetilde{\mathbf{N}})_{:,i}\|_2 + \beta\|\mathbf{P}_{\widetilde{\mathcal{L}}^\perp}\widetilde{\mathbf{N}}_{:,j}\|_2$$
$$\quad + \beta\frac{18\gamma n_2\eta_{\mathbf{N}}}{\sigma_r(\check{\mathbf{L}})}(\|\widetilde{\mathbf{L}}_{:,j}\|_2 + \|\widetilde{\mathbf{N}}_{:,j}\|_2)$$
$$\stackrel{(v)}{=} \frac{\tau_1}{4}\|(\widetilde{\mathbf{C}}+\widetilde{\mathbf{N}})_{:,i}\|_2 + \beta\|\mathbf{P}_{\widetilde{\mathcal{L}}^\perp}\widetilde{\mathbf{N}}_{:,j}\|_2$$
$$\quad + \beta\tau_3(\|\widetilde{\mathbf{L}}_{:,j}\|_2 + \|\widetilde{\mathbf{N}}_{:,j}\|_2)$$
$$\stackrel{(vi)}{\geq} \tau_3\|(\widetilde{\mathbf{C}}+\widetilde{\mathbf{N}})_{:,i}\|_2 + \beta(\tau_3\|(\widetilde{\mathbf{L}}+\widetilde{\mathbf{N}})_{:,j}\|_2 + \|\mathbf{P}_{\widetilde{\mathcal{L}}^\perp}\widetilde{\mathbf{N}}_{:,j}\|_2), \quad (41)$$

where $(i)$ is from (40), $(ii)$ is from (32), $(iii)$ is from ($\widetilde{d}$4), $(iv)$ is from (32) and fact that $\max_{j \in [n_2]} \|\widetilde{\mathbf{L}}_{:,j}\|_2 \leq \sigma_1(\widetilde{\mathbf{L}})$, $(v)$ is from (36), and $(vi)$ is from (37). Combining (38), (39) and (41), we have

$$\|\mathbf{P}_{\widehat{\mathcal{L}}_{(1)}^\perp}(\widetilde{\mathbf{C}}+\widetilde{\mathbf{N}})_{:,j}\|_2 > \beta\left(\|\mathbf{P}_{\widetilde{\mathcal{L}}^\perp}\widetilde{\mathbf{N}}_{:,j}\|_2 + \tau_3\|(\widetilde{\mathbf{L}}+\widetilde{\mathbf{N}})_{:,j}\|_2\right)$$
$$\geq \beta\|\mathbf{P}_{\widehat{\mathcal{L}}_{(1)}^\perp}(\widetilde{\mathbf{L}}+\widetilde{\mathbf{N}})_{:,j}\|_2.$$

Then Claim (II) of Theorem II.1 is verified.

### C. Proof of Lemma III.3

We first show that when $q$ satisfies (8), the random projection via $\boldsymbol{\Psi} \in \mathbb{R}^{q \times m}$ approximately preserves (24) with high probability. This is formalized in the following lemma.

**Lemma VI.2.** *Suppose $\boldsymbol{\Psi} \in \mathbb{R}^{q \times m}$ is drawn from a distribution satisfying the distributional JL property (4) with $q$ satisfying (8), and (24) holds. Given $\delta \in (0,1)$, then for any $i \in \mathcal{I}_{\mathbf{C}}$ and $j \in \mathcal{I}_{\mathbf{L}}$, with probability at least $1-\delta$, we have*

$$\|\boldsymbol{\Psi}\mathbf{P}_{\widehat{\mathcal{L}}_{(1)}^\perp}(\widetilde{\mathbf{C}}+\widetilde{\mathbf{N}})_{:,i}\|_2 > \frac{\sqrt{3}}{3}\beta\|\boldsymbol{\Psi}\mathbf{P}_{\widehat{\mathcal{L}}_{(1)}^\perp}(\widetilde{\mathbf{L}}+\widetilde{\mathbf{N}})_{:,j}\|_2. \quad (42)$$

*Proof.* For any $\varepsilon \in (0,1)$, if $\boldsymbol{\Psi}$ is as specified and (24) holds, then with probability at least $1-\delta$, we have

$$\|\boldsymbol{\Psi}\mathbf{P}_{\widehat{\mathcal{L}}_{(1)}^\perp}(\widetilde{\mathbf{C}}+\widetilde{\mathbf{N}})_{:,i}\|_2 \geq \sqrt{\frac{1-\varepsilon}{1+\varepsilon}}\|\mathbf{P}_{\widehat{\mathcal{L}}_{(1)}^\perp}(\widetilde{\mathbf{C}}+\widetilde{\mathbf{N}})_{:,j}\|_2$$
$$> \sqrt{\frac{1-\varepsilon}{1+\varepsilon}}\beta\|\mathbf{P}_{\widehat{\mathcal{L}}_{(1)}^\perp}(\widetilde{\mathbf{L}}+\widetilde{\mathbf{N}})_{:,j}\|_2$$
$$\geq \sqrt{\frac{1-\varepsilon}{1+\varepsilon}}\beta\|\boldsymbol{\Psi}\mathbf{P}_{\widehat{\mathcal{L}}_{(1)}^\perp}(\widetilde{\mathbf{L}}+\widetilde{\mathbf{N}})_{:,j}\|_2.$$

Then (42) holds if we take $\varepsilon = \sqrt{2}/4$. $\square$

Now it is straightforward to see that if $\beta > \sqrt{3}$ and we choose some $\varepsilon_2$ that satisfies $\max_{j \in \mathcal{I}_{\mathbf{L}}} \|\boldsymbol{\Psi}\mathbf{P}_{\widehat{\mathcal{L}}_{(1)}^\perp}(\widetilde{\mathbf{L}}+\widetilde{\mathbf{N}})_{:,j}\|_2 < \varepsilon_2 < \min_{i \in \mathcal{I}_{\mathbf{C}}} \|\boldsymbol{\Psi}\mathbf{P}_{\widehat{\mathcal{L}}_{(1)}^\perp}(\widetilde{\mathbf{C}}+\widetilde{\mathbf{N}})_{:,i}\|_2$, then we have $\widehat{\mathcal{I}}_{\mathbf{C}} = \mathcal{I}_{\widetilde{\mathbf{C}}}$.

### D. Proof of Theorem II.2

We only need to bound the constants $\eta_{\mathbf{N}}$, which further implies the bound of $\alpha$. The rest of the proof is identical to that of Theorem II.1. Since $\mathbf{N}$ has i.i.d. $\mathcal{N}(0, \sigma^2)$ entries, then for any $i \in [n_2]$, $\|\mathbf{N}_{:,i}\|_2^2/\sigma^2 = \sum_{j=1}^{n_1}(N_{ji}/\sigma)^2$ has chi-square distribution $\chi^2_{n_1}$ with $n_1$ degree of freedom. Given $t \in (0,1)$, we have the following tail bounds [40],

$$\mathcal{P}\left(\|\mathbf{N}_{:,i}\|_2^2 \geq \sigma^2 n_1(1+t)\right) \leq e^{-n_1 t^2/8} \text{ and}$$
$$\mathcal{P}\left(\|\mathbf{N}_{:,i}\|_2^2 \leq \sigma^2 n_1(1-t)\right) \leq e^{-n_1 t^2/8}.$$

Let $t = \sqrt{s/n_1}$ for some $s \in (0, n_1)$, then we have

$$\mathcal{P}\left(\|\mathbf{N}_{:,i}\|_2 \geq \sigma\sqrt{n_1 + \sqrt{n_1 s}}\right) \leq e^{-s/8} \text{ and}$$
$$\mathcal{P}\left(\|\mathbf{N}_{:,i}\|_2 \leq \sigma\sqrt{n_1 - \sqrt{n_1 s}}\right) \leq e^{-s/8}.$$

By union bound, we further have

$$\mathcal{P}\left(\max_{i \in [n_2]} \|\mathbf{N}_{:,i}\|_2 \leq \sigma\sqrt{n_1 - \sqrt{n_1 s}}\right) \leq n_2 e^{-s/8} \text{ and}$$
$$\mathcal{P}\left(\max_{i \in [n_2]} \|\mathbf{N}_{:,i}\|_2 \geq \sigma\sqrt{n_1 + \sqrt{n_1 s}}\right) \leq n_2 e^{-s/8},$$

Let $\delta = 2n_2 e^{-s/8} \in (0,1)$ and apply the union bound, then with probability at least $1-\delta$, we have

$$C_1\sigma \leq \eta_{\mathbf{N}} \leq C_2\sigma, \quad (43)$$

where $C_1$ and $C_2$ are specified as in Theorem II.2.

Combining (13) and (43), we have $\alpha$ satisfies (11). Finally, the overall results of Theorem II.2 hold via the union bound.

*E. Proof of Theorem II.3*

We follow the idea of the proof for Theorem II.1 by providing the proof sketch for Theorem II.3, which is formalized by the intermediate results Lemma VI.3 and Lemma VI.4. The proofs of the lemmata are provided later.

For notional convenience, we introduce:

$$\widetilde{\mathbf{M}} = \mathbf{\Phi}\mathbf{M}, \ \widetilde{\mathbf{L}} = \mathbf{\Phi}\mathbf{L}, \ \widetilde{\mathbf{C}} = \mathbf{\Phi}\mathbf{C}, \text{ and } \widetilde{\mathbf{C}}_{\mathbf{\Omega}} = \mathbf{\Phi}\mathbf{P}_{\mathbf{\Omega}}(\mathbf{C}).$$

We start with showing that the analogous structural conditions for $\widetilde{\mathbf{L}}$ and $\widetilde{\mathbf{C}}$ also hold for $\widetilde{\mathbf{M}}$ provided that the row sampling parameter $\gamma_1$ is sufficiently large. This is stated in Lemma VI.3, and we provide the proof in Section VI-E1.

**Lemma VI.3.** *Suppose matrices $\mathbf{L}, \mathbf{C} \in \mathbb{R}^{n_1 \times n_2}$ satisfy the structural conditions (g1)-(g4) with $p$ satisfying (14). Given $\delta \in (0,1)$, further suppose $\mathbf{\Phi} \in \mathbb{R}^{m \times n_1}$ is a row sampling matrix with the sampling parameter $\gamma_1$ satisfying (16). Then, with probability at least $1-\delta$, the components $\widetilde{\mathbf{L}}$ and $\widetilde{\mathbf{C}}$ satisfy*

(g̃1) $\mathrm{rank}(\widetilde{\mathbf{L}}) = r$,
(g̃2) $\widetilde{\mathbf{L}}$ has $n_{\mathbf{L}}$ nonzero columns,
(g̃3) $\widetilde{\mathbf{L}}$ satisfies the row and column incoherence property with parameters $\mu_{\widetilde{\mathbf{U}}} = 9\mu_{\mathbf{U}}$ and $\mu_{\widetilde{\mathbf{V}}} = \mu_{\mathbf{V}}$ respectively, and
(g̃4) $\mathcal{I}_{\widetilde{\mathbf{C}}} \triangleq \left\{ j \in [n_2] : \|(\mathbf{P}_{\widetilde{\mathcal{L}}^\perp}(\widetilde{\mathbf{C}}_{\mathbf{\Omega}}))_{:,j}\|_2 > 0 \right\} = \mathcal{I}_{\mathbf{C}}$, where $\widetilde{\mathcal{L}}$ denotes the subspace spanned by columns of $\widetilde{\mathbf{L}}$, and $\mathbf{P}_{\widetilde{\mathcal{L}}^\perp}$ is the orthogonal projection onto the orthogonal complement of $\widetilde{\mathbf{L}}$.

*Let $\mu_{\widetilde{\mathbf{L}}} = \max(\mu_{\widetilde{\mathbf{U}}}, \mu_{\widetilde{\mathbf{V}}})$. Simultaneously, we have*

(r1) $\mathbf{\Phi}$ has $(1/2)\gamma_1 n_1 \leq m \leq (3/2)\gamma_1 n_1$ rows,
(r2) each column of $\widetilde{\mathbf{L}}_{\mathbf{\Omega}}$ has at least $4r\mu_{\mathbf{L}} \log(2r)$ observed entries, and
(r3) $p$ satisfies $p \geq C_p \frac{\mu_{\widetilde{\mathbf{L}}}^2 r^2 \log^3(4n_{\mathbf{L}})}{m}$.

The next result guarantees that when the column sampling parameter $\gamma_2$ is sufficiently large, exact outlier detection may be achieved. This is formalized in Lemma VI.4, and we provide the proof in Section VI-E2.

**Lemma VI.4.** *Suppose $\widetilde{\mathbf{L}}, \widetilde{\mathbf{C}} \in \mathbb{R}^{m \times n_2}$ satisfy the structural conditions (g̃1)-(g̃4) with $k$ satisfying (15), and (r1)-(r3) hold. Given $\delta \in (0,1)$, suppose the column sampling parameter $\gamma_2$ satisfies (17) and $\lambda$ satisfies (18), then the following claims hold simultaneously with probability at least $1-\delta$:*

(I) $\widehat{\mathcal{I}}_{\mathbf{C}} = \mathcal{I}_{\widetilde{\mathbf{C}}}$, *i.e. the estimate of the outlier identities is exact, and*
(II) *the total number of measurements collected is no greater than $\frac{3}{2}p\gamma_1 n_1 n_2$.*

The overall results of Theorem II.3 follow by combining two intermediate lemmata via the union bound.

*1) Proof of Lemma VI.3:* Part 1: We first verify (r1)-(r3). We show that when $\gamma_1$ satisfies

$$\gamma_1 \geq \max \left\{ \frac{2r\mu_{\mathbf{U}} \log(2r)}{n_1 p}, \frac{8 \log \frac{4n_{\mathbf{L}}}{\delta}}{n_1 p}, \frac{10r\mu_{\mathbf{U}} \log \frac{4r}{\delta}}{n_1} \right\}, \quad (44)$$

then with high probability, we have

(h1) $\mathbf{\Phi}$ has $(1/2)\gamma_1 n_1 \leq m \leq (3/2)\gamma_1 n_1$ rows,
(h2) each column of $\widetilde{\mathbf{L}}$ has at least $4r\mu_{\mathbf{U}} \log(2r)$ observed entries,
(h3) $\sigma_1^2(\mathbf{\Phi}\mathbf{U}) \leq (3/2)\gamma_1$, and
(h4) $\sigma_r^2(\mathbf{\Phi}\mathbf{U}) \geq (1/2)\gamma_1$.

Let $\mathcal{E}_1, \ldots, \mathcal{E}_4$ denote the events that (h1)-(h4) hold respectively. Then $\Pr\left(\left\{\bigcap_{i=1}^4 \mathcal{E}_i\right\}^c\right) \leq \sum_{i=1}^4 \Pr(\mathcal{E}_i^c)$, and we consider each term in the sum as follows.

First, since $m$ is a Binomial$(n_1, \gamma_1)$ random variable, we bound its tails using [41, Theorem 2.3 (b-c)]. This gives that $\Pr(m > 3\gamma_1 n_1/2) \leq \exp(-3\gamma_1 n_1/28)$ and $\Pr(m < \gamma_1 n_1/2) \leq \exp(-\gamma_1 n_1/8)$. By union bound, we obtain that $\Pr(\mathcal{E}_1^c) \leq \exp(-3\gamma_1 n_1/28) + \exp(-\gamma_1 n_1/8)$.

Next, the number of observed entries in each column is a Binomial$(n_1, p\gamma_1)$. Leveraging the result in [42], we have that the minimum number of observed entries requested in the non-zero column of $\mathbf{L}$ is $4r\mu_{\mathbf{U}} \log(2r)$ for that column to be recovered correctly with probability 1. Therefore, we need $\gamma_1 n_1 p/2 \geq 4r\mu_{\mathbf{U}} \log(2r)$, which is equivalent to

$$\gamma_1 \geq \frac{2r\mu_{\mathbf{U}} \log(2r)}{n_1 p}. \quad (45)$$

Further, by the union bound, we have

$$\Pr(\cup_{j \in \mathcal{I}_{\mathbf{L}}}\{|\mathcal{I}_j| \leq \gamma_1 n_1 p/2\}) \leq \sum_{j \in \mathcal{I}_{\mathbf{L}}} \Pr(|\mathcal{I}_j| \leq \gamma_1 n_1 p/2)$$
$$\leq n_{\mathbf{L}} \exp\{-\frac{\gamma_1 n_1 p}{8}\},$$

Then we have $\Pr(\mathcal{E}_2^c) \leq n_{\mathbf{L}} \exp\{-\frac{\gamma_1 n_1 p}{8}\}$.

Finally, applying [43, Corollary 5.2], we obtain

$$\Pr(\mathcal{E}_3^c) = \Pr\left(\sigma_1^2(\mathbf{\Phi}\mathbf{U}) \geq 3\gamma_1/2\right) \leq r \cdot (9/10)^{\frac{\gamma_1 n_1}{r\mu_{\mathbf{U}}}} \text{ and}$$
$$\Pr(\mathcal{E}_4^c) = \Pr\left(\sigma_r^2(\mathbf{\Phi}\mathbf{U}) \leq \gamma_1/2\right) \leq r \cdot (9/10)^{\frac{\gamma_1 n_1}{r\mu_{\mathbf{U}}}}.$$

Putting these results together, we have

$$\Pr\left(\left\{\bigcap_{i=1}^4 \mathcal{E}_i\right\}^c\right) \leq \exp(-3\gamma_1 n_1/28) + \exp(-\gamma_1 n_1/8)$$
$$+ n_{\mathbf{L}} \exp\left\{-\frac{\gamma_1 n_1 p}{8}\right\} + 2r \cdot (9/10)^{\frac{\gamma_1 n_1}{r\mu_{\mathbf{U}}}}. \quad (46)$$

The R.H.S. of (46) is upper bounded by $\delta$ given that each term in the sum is no larger than $\delta/4$. This requires

$$\gamma_1 \geq \left\{ \frac{8 \log \frac{4n_{\mathbf{L}}}{\delta}}{n_1 p}, \frac{10r\mu_{\mathbf{U}} \log \frac{4r}{\delta}}{n_1} \right\}. \quad (47)$$

Combing (45) and (47), we have (44).

The condition (r1) and (r2) follow directly from (h1) and (h2). Next, we verify (r3). Given $m \geq \gamma_1 n_1/2$, for successful outlier identification via matrix completion [14], we require

$$p \geq C_p \frac{81\mu_{\mathbf{L}}^2 r^2 \log^3(4n_{\mathbf{L}})}{\gamma_1 n_1/2} \geq C_p \frac{\mu_{\widetilde{\mathbf{L}}}^2 r^2 \log^3(4n_{\mathbf{L}})}{m}, \quad (48)$$

where $\mu_{\widetilde{\mathbf{L}}} = 9\mu_{\mathbf{L}}$ and (r3) follows. This requires

$$\gamma_1 \geq C_p \frac{162\mu_{\mathbf{L}}^2 r^2 \log^3(4n_{\mathbf{L}})}{pn_1} = \frac{162p_l}{p}. \quad (49)$$

Combining (44) and (49), we have the bound (16) for $\gamma_1$.

Part 2: Next, we show that (**g̃1**)-(**g̃4**) follow directly when (**g1**)-(**g4**), (**h3**) and (**h4**) hold. The condition (**g̃1**) and the first part of (**g̃3**) ($\mu_{\widetilde{U}} = 9\mu_{U}$) follow directly from (**h3**) and (**h4**) with the argument in [15] (Lemma III.2). The second part of (**g̃3**) ($\mu_{\widetilde{V}} = \mu_{V}$) is based on the fact that $\widetilde{L}$ is a row submatrix of $L$, where $\widetilde{L}$ and $L$ share an identical row spaces since $\text{rank}(\widetilde{L}) = \text{rank}(L) = r$. Therefore, they have the same column incoherence parameter (two right singular vectors are only of a difference of rotation). The condition (**g̃2**) is a direct result from (**g2**). The condition (**g̃4**) is a direct result from (**g4**), since for any row submatrix $\widetilde{C}_{\Omega}$ of $C_{\Omega}$, where $\Phi$ has $m \geq (1/2)\gamma_1 n_1 \geq r\mu_U \log(2r)/p$ rows, the orthogonal projection of nonzero columns of $\Phi P_{\Omega} C$ onto the orthogonal complement of column space of $\widetilde{L}$ is not zero.

*2) Proof of Lemma VI.4:* For notational convenience, we introduce
$$\check{M} = \widetilde{M}S, \quad \check{L} = \widetilde{L}S, \quad \text{and} \quad \check{C} = \widetilde{C}S.$$

Using a straightforward modification of the intermediate result of Lemma III.2 in [15] again with a slightly different constant ($\delta/9$ rather than $\delta/5$), we have if $\gamma_2$ satisfies
$$\gamma_2 \geq \max\left\{\frac{200\log(\frac{9}{\delta})}{n_L}, \frac{10r\mu_V \log(\frac{9r}{\delta})}{n_L}, \frac{200\log(\frac{9}{\delta})}{k_u}\right\}, \quad (50)$$

then with probability at least $1 - \frac{8}{9}\delta$, we have
(**b1**) $S$ has $(1/2)\gamma_2 n_2 \leq |\mathcal{S}_2| \leq (3/2)\gamma_2 n_2$ columns,
(**b2**) $\check{L}$ has $(1/2)\gamma_2 n_L \leq n_{\check{L}} \leq (3/2)\gamma_2 n_L$ nonzero columns,
(**b3**) $\check{C}$ has at most $(3/2)\gamma_2 k_u$ nonzero columns,
(**b4**) $\sigma_1^2(\widetilde{V}^* S) \leq (3/2)\gamma_2$, and
(**b5**) $\sigma_r^2(\widetilde{V}^* S) \geq (1/2)\gamma_2$.

Moreover, if (**b1**)-(**b5**) hold, we have the following structural properties of $\check{L}$ and $\check{C}$:
(**ǧ1**) $\text{rank}(\check{L}) = r$,
(**ǧ2**) $\check{L}$ has $n_{\check{L}} \leq \frac{3}{2}\gamma_2 n_L$ nonzero columns,
(**ǧ3**) $\check{L}$ satisfies the *row and column incoherence properties* with parameters $\mu_{\check{U}} = 9\mu_U$ and $\mu_{\check{V}} = 9\mu_V$ respectively, and
(**ǧ4**) $\mathcal{I}_{\check{C}} \triangleq \{j : \|P_{\check{\mathcal{L}}^{\perp}} \check{C}_{:,j}\|_2 > 0\}$ with $|\mathcal{I}_{\check{C}}| = \check{k}$, where $\check{\mathcal{L}}$ denotes the subspace spanned by columns of $\check{L}$ and $P_{\check{\mathcal{L}}^{\perp}}$ is the orthogonal projection operation onto the orthogonal complement of $\check{\mathcal{L}}$.

Next, we show that MP in Step 1 of RACOS-I succeeds with high probability under proper conditions. Let $\mu_{\check{L}} = \max(\mu_{\check{U}}, \mu_{\check{V}})$, $\check{k}$ be the number of outliers in $\check{M}$, and $\check{n}_2 = |\mathcal{S}_2|$ be the number columns of $\check{M}$. We formalize the result in Lemma VI.5 without proof.

**Lemma VI.5** (Adapted from Theorem 1 in [14]). *Suppose $\check{L}$ and $\check{C}$ satisfy the structural conditions (**ǧ1**)-(**ǧ4**) with $n_{\check{L}} \geq m \geq 32$. If $(r, \check{k}, p)$ satisfies (48) and*
$$\check{k} \leq \frac{p^2 \check{n}_2}{p^2 + C_{\check{k}}(1 + \frac{\mu_{\check{L}}r}{p\sqrt{m}})\mu_{\check{L}}^3 r^3 \log^6(4n_{\check{L}})}, \quad (51)$$

*, where $C_{\check{k}}$ is a constant, and $\lambda$ satisfies*
$$\lambda = \frac{1}{48}\sqrt{\frac{p}{kr\mu_{\check{L}} \log^2(4n_{\check{L}})}},$$

*then MP returns $\{\widehat{L}_{(1)}, \widehat{C}_{(1)}\}$ such that $\widehat{L}_{(1)}$ has the same column space with $\check{L}$, and $\widehat{C}_{(1)}$ has the same column support with $\check{C}$, with probability at least $1 - \check{C}_{\gamma_2}\check{n}_2^{-5}$ for some positive constant $\check{C}_{\gamma_2}$.*

We start by verifying that $p$ and $\check{k}$ satisfy the required bound for the data matrix $\check{M}$ in Lemma VI.5. Given $p$ satisfying the bound in (**r3**) of Lemma VI.3 and $n_L \geq n_{\check{L}}$, we have that
$$p \geq C_p \frac{\mu_{\check{L}}^2 r^2 \log^3(4n_{\check{L}})}{m}, \quad (52)$$

where $\mu_{\check{L}} = \mu_{\widetilde{L}} = 9\mu_L$.

For $\check{k}$, when (50) and $\frac{200}{\gamma_2}\log(\frac{9}{\delta}) \leq k \leq k_u$ hold, with probability at least $1 - \delta/9$, we have
$$\check{k} \leq \frac{3}{2}\gamma_2 k_u = \frac{p^2 \gamma_2 n_2/2}{p^2 + C_k(1 + \frac{3\sqrt{6}\mu_L r}{p\sqrt{n_1}})n_L^3 r^3 \log^6(4n_L)}$$
$$\leq \frac{p^2 \gamma_2 n_2/2}{p^2 + C_{\check{k}}(1 + \frac{9\mu_L r}{p\sqrt{\frac{3}{2}\gamma_1 n_1}})(9\mu_L)^3 r^3 \log^6(4n_L)}$$
$$\leq \frac{p^2 \check{n}_2}{p^2 + C_{\check{k}}(1 + \frac{\mu_{\check{L}}r}{p\sqrt{m}})\mu_{\check{L}}^3 r^3 \log^6(4n_{\check{L}})},$$

where $C_{\check{k}} = C_k/729$. When $k < \frac{200}{\gamma_2}\log(\frac{9}{\delta})$, we have from Lemma 4.1 in [16] that $\Pr(\check{k}_2 \leq t) \geq \Pr(\check{k}_1 \leq t)$ for any $t \in [0, \infty)$, where $\check{k}_1$ and $\check{k}_2$ are Hypergeometric random variables with distributions $\text{Hyp}(n_2, \check{n}_2, k_1)$ and $\text{Hyp}(n_2, \check{n}_2, k_2)$ respectively. This implies with probability at least $1 - 9/\delta$, we have
$$\check{k} \leq \frac{3}{2}\gamma_2 k_u \leq \frac{p^2 \check{n}_2}{p^2 + C_{\check{k}}(1 + \frac{\mu_{\check{L}}r}{p\sqrt{m}})\mu_{\check{L}}^3 r^3 \log^6(4n_{\check{L}})}. \quad (53)$$

This also verifies (**b3**).

From $n_L \geq \frac{\delta e^{8p}}{4}$ and $\gamma_1 \geq \frac{8\log\frac{4n_L}{\delta}}{n_1 p}$, if
$$\gamma_2 \geq \frac{3\gamma_1 n_1}{n_L}, \quad (54)$$

we have $n_{\check{L}} \geq m \geq 32$. Exact outlier identification is achievable with high probability when (52), (53), and (51) hold.

Let $\check{C}_{\gamma_2} \check{n}_2^{-5} \leq \check{C}_{\gamma_2}\left(\frac{3}{2}\gamma_2 n_2\right)^{-5} \leq \delta/9$, then we have that
$$\gamma_2 \geq \frac{3}{2n_2}\left(\frac{9\check{C}_{\gamma_2}}{\delta}\right)^{\frac{1}{5}} = \frac{C_{\gamma_2}(\frac{1}{\delta})^{\frac{1}{5}}}{n_2}, \quad (55)$$

where $C_{\gamma_2} = \frac{3}{2}(9\check{C}_{\gamma_2})^{\frac{1}{5}}$. Combining (50), (54) and (55), we obtain the bound (17) for $\gamma_2$.

Note that $\widehat{\mathcal{L}}_{(1)} = \check{\mathcal{L}} = \widetilde{\mathcal{L}}$ and the number of observed entries for each non-zero column of $\widetilde{L}$ is at least $4r\mu_U \log(2r)$ from (**r2**) of Lemma VI.3. Then we have that for any $j \in \mathcal{I}_L$, $\|P_{\widehat{L}_{\mathcal{I}_j}^{\perp}} \widetilde{L}_{\mathcal{I}_j,j}\| = 0$ with probability 1 from [42] (Theorem 1), and for any $j \in \mathcal{I}_C$, $\|P_{\widehat{L}^{\perp}} \widetilde{C}_{\mathcal{I}_j,j}\| > 0$ from (**g̃4**). Therefore, we have $\widehat{\mathcal{I}}_C = \mathcal{I}_{\widetilde{C}}$, thus Claim (I) follows. Claim (II) holds directly from (**r1**). Finally, the overall result holds with probability at least $1 - \delta$ using the union bound.

### F. Proof of Theorem II.4

The analysis of Theorem II.4 is analogous to that of Theorem II.3. For completeness, we provide the intermediate results here. We first show that the structural conditions for $\widetilde{\mathbf{L}}$ and $\widetilde{\mathbf{C}}$ provided that the row sampling parameter $\gamma_1$ is sufficiently large. This is formalized in Lemma VI.6, and we provide the proof in Section VI-F1.

**Lemma VI.6.** *Suppose* $\mathbf{L}, \mathbf{C} \in \mathbb{R}^{n_1 \times n_2}$ *satisfy the structural conditions* **(g1)**-**(g4)** *with $p$ satisfying* (19). *Given* $\delta \in (0,1)$, *suppose* $\mathbf{\Phi} \in \mathbb{R}^{m \times n_1}$ *is a row sampling matrix with the sampling parameter $\gamma_1$ satisfying* (16). *Then, with probability at least $1 - \delta$, the components $\widetilde{\mathbf{L}}$ and $\widetilde{\mathbf{C}}$ satisfy*

- **(g̃1)** $\mathrm{rank}(\widetilde{\mathbf{L}}) = r$,
- **(g̃2)** $\widetilde{\mathbf{L}}$ *has $n_{\mathbf{L}}$ nonzero columns*,
- **(g̃3)** $\widetilde{\mathbf{L}}$ *satisfies the* row and column incoherence property *with parameters $\mu_{\widetilde{\mathbf{U}}} = 9\mu_{\mathbf{U}}$ and $\mu_{\widetilde{\mathbf{V}}} = \mu_{\mathbf{V}}$ respectively, and*
- **(g̃4)** $\mathcal{I}_{\widetilde{\mathbf{C}}} \triangleq \left\{ j \in [n_2] : \|(\mathbf{P}_{\widetilde{\mathcal{L}}^\perp}(\widetilde{\mathbf{C}}_{\mathbf{\Omega}}))_{:,j}\|_2 > 0 \right\} = \mathcal{I}_{\mathbf{C}}$, *where $\widetilde{\mathcal{L}}$ denotes the subspace spanned by columns of $\widetilde{\mathbf{L}}$, and $\mathbf{P}_{\widetilde{\mathcal{L}}^\perp}$ is the orthogonal projection onto the orthogonal complement of $\widetilde{\mathbf{L}}$ in $\mathbb{R}^m$.*

*Let $\mu_{\widetilde{\mathbf{L}}} = \max(\mu_{\widetilde{\mathbf{U}}}, \ \mu_{\widetilde{\mathbf{V}}})$. Simultaneously, we have*

- **(r1)** $\mathbf{\Phi}$ *has $(1/2)\gamma_1 n_1 \leq m \leq (3/2)\gamma_1 n_1$ rows*,
- **(r2)** *each column of $\widetilde{\mathbf{L}}_{\mathbf{\Omega}}$ has at least $4r\mu_{\mathbf{L}} \log(2r)$ observed entries, and*
- **(r3)** $p$ *satisfies* $p \geq C_p \frac{\mu_{\widetilde{\mathbf{L}}}^2 r^2 \log^3(4n_{\mathbf{L}})}{m}$.

The next result guarantees that when the column sampling parameter $\gamma_2$ is sufficiently large, exact outlier detection may be achieved. This is formalized in Lemma VI.7, and we provide the proof in Section VI-F2.

**Lemma VI.7.** *Suppose* $\widetilde{\mathbf{L}}, \widetilde{\mathbf{C}} \in \mathbb{R}^{m \times n_2}$ *satisfy the conditions* **(g̃1)**-**(g̃4)** *with $k$ satisfying* (20), *and* **(r1)**-**(r3)** *hold. Given* $\delta \in (0,1)$, *suppose the column sampling parameter $\gamma_2$ satisfies* (17) *and $\lambda$ satisfies* (21), *then the following hold simultaneously with probability at least $1 - \delta$:*

(I) $\widehat{\mathcal{I}}_{\mathbf{C}} = \mathcal{I}_{\widetilde{\mathbf{C}}}$, *i.e. the estimate of the outlier identities is exact, and*

(II) *the total number of measurements collected is no greater than $\frac{3}{2} p \gamma_1 n_1 n_2$.*

The overall result of Theorem II.4 follows by combining two intermediate results via the union bound.

*1) Proof of Lemma VI.6:* The analysis follows directly from that of Lemma VI.3, except for **(r3)**. Given $m \geq \gamma_1 n_1 / 2$, for successful outlier identification via MP [14], we need $p$ to satisfy

$$p \geq C_p \left(1 + \frac{1}{\varphi}\right) \frac{9\mu_{\mathbf{L}} r \log^2(2n_2)}{\gamma_1 n_1 / 2}$$
$$\geq C_p \left(1 + \frac{1}{\varphi}\right) \frac{\mu_{\widetilde{\mathbf{L}}} r \log^2(2\check{n}_2)}{m}, \quad (56)$$

which is implied by (19), where $\mu_{\widetilde{\mathbf{L}}} = 9\mu_{\mathbf{L}}$ and **(r3)** follows. This requires

$$\gamma_1 \geq C_p \left(1 + \frac{1}{\varphi}\right) \frac{18 \mu_{\mathbf{L}} r \log^2(2n_2)}{p n_1} = \frac{18 p_l}{p}. \quad (57)$$

Combining (44) and (56), we have the bound (16) for $\gamma_1$.

*2) Proof of Lemma VI.7:* The analysis follows directly from that of Lemma VI.4, except the bound of $k$. For successful outlier identification via MP [14], we need $\check{k}$ to satisfy

$$\check{k} \leq C_{\check{k}} \frac{\varphi}{1 + \varphi \sqrt{\varphi}} \frac{p n_{\widetilde{\mathbf{L}}}}{\mu_{\widetilde{\mathbf{L}}}^{3/2} r^{3/2} \log^3(2n_2)},$$

where $C_{\check{k}}$ is a constant, which holds since $\check{k} \leq \frac{3}{2} \gamma_2 k_u$ and $k_u$ satisfies (20).

### G. Proof of Lemma VI.1

We leverage the intermediate result of the proof of Theorem 2 in [9], which provides the estimation error bounds of both low-rank and outlier components w.r.t. the noise in term of the Frobenius norm. However, we are interested in the the estimation error bound of the low-rank component in terms of the spectral norm, which are the main technical differences in our proof here.

To start with, we define two operators that return the subgradient of $\|\mathbf{L}\|_*$ and $\|\mathbf{C}\|_{1,2}$ in the following lemma. Recall that $\mathbf{P}_{\mathcal{L}}(\cdot)$ is an orthogonal projection operator that project a matrix to the column space $\mathcal{L}$ of $\mathbf{L}$, and $\mathbf{P}_{\mathcal{I}_{\mathbf{C}}}(\cdot)$ is a projection operator that leave columns in the support set $\mathcal{I}_{\mathbf{C}}$ unchanged and set the other columns to be $\mathbf{0}$.

**Definition VI.1.** *Let* $\mathbf{M} = \mathbf{L}' + \mathbf{C}'$, *where* $\mathbf{P}_{\mathcal{L}}(\mathbf{L}') = \mathbf{L}'$ *and* $\mathbf{P}_{\mathcal{I}_{\mathbf{C}}}(\mathbf{C}') = \mathbf{C}'$. *Given the compact SVD of* $\mathbf{L}'$ *as* $\mathbf{L}' = \mathbf{U}' \mathbf{\Sigma}' \mathbf{V}'^T$ *and the column support of* $\mathbf{C}'$ *as* $\mathcal{I}'_{\mathbf{C}}$, *we define the following:*

$\mathfrak{R}(\mathbf{L}') \triangleq \mathbf{U}'\mathbf{V}'^T$;

$\mathfrak{G}(\mathbf{C}') \triangleq \left\{ \mathbf{H} \in \mathbb{R}^{n_1 \times n_2} \big| \mathbf{P}_{\mathcal{I}_{\mathbf{C}}^c} = \mathbf{0}; \ \forall i \in \mathcal{I}'_{\mathbf{C}} \subseteq \mathcal{I}_{\mathbf{C}}, \ \mathbf{H}_{:,i} = \mathbf{C}'_{:,i} / \|\mathbf{C}'_{:,i}\|_2; \ \forall i \in \mathcal{I}_{\mathbf{C}} \cap (\mathcal{I}'_{\mathbf{C}})^c, \ \|\mathbf{H}_{:,i}\|_2 < 1 \right\}.$

Consider the noisy OP problem (29). Let $\check{\mathbf{M}}_0 = \check{\mathbf{L}}_1 + \check{\mathbf{C}}_1$, where $\mathbf{P}_{\check{\mathcal{L}}}(\check{\mathbf{L}}_1) = \check{\mathbf{L}}_1$ and $\mathbf{P}_{\mathcal{I}_{\check{\mathbf{C}}}}(\check{\mathbf{C}}_1) = \check{\mathbf{C}}_1$. For $\check{\mathbf{L}}_1 = \check{\mathbf{U}}_1 \check{\mathbf{\Sigma}}_1 \check{\mathbf{V}}_1^T$ and $\check{\mathbf{L}} = \check{\mathbf{U}} \check{\mathbf{\Sigma}} \check{\mathbf{V}}^T$, there exists an orthonormal matrix $\overline{\mathbf{V}} \in \mathbb{R}^{r \times \check{n}_2}$, such that $\check{\mathbf{U}}_1 \check{\mathbf{V}}_1^T = \check{\mathbf{U}} \overline{\mathbf{V}}^T$. Further let $\mathbf{P}_{\mathcal{T}(\check{\mathbf{L}}_1)}$ be the projection onto the space spanned by $\check{\mathbf{U}}_1$ and $\check{\mathbf{V}}_1$, which is given by $\mathbf{P}_{\mathcal{T}(\check{\mathbf{L}}_1)} = \mathbf{P}_{\check{\mathbf{U}}_1} + \mathbf{P}_{\check{\mathbf{V}}_1} - \mathbf{P}_{\check{\mathbf{U}}_1} \mathbf{P}_{\check{\mathbf{V}}_1}$, $\check{\mathbf{N}}_{\mathbf{L}} = \widehat{\mathbf{L}} - \check{\mathbf{L}}_1$, and $\check{\mathbf{N}}_{\mathbf{C}} = \widehat{\mathbf{C}} - \check{\mathbf{C}}_1$, thus $\check{\mathbf{N}} = \check{\mathbf{N}}_{\mathbf{L}} + \check{\mathbf{N}}_{\mathbf{C}}$. Define $\check{\mathbf{N}}_{\mathbf{L}}^+ = \check{\mathbf{N}}_{\mathbf{L}} - \mathbf{P}_{\mathcal{I}_{\check{\mathbf{C}}}} \mathbf{P}_{\check{\mathcal{L}}}(\check{\mathbf{N}}_{\mathbf{L}})$, $\check{\mathbf{N}}_{\mathbf{C}}^+ = \check{\mathbf{N}}_{\mathbf{C}} - \mathbf{P}_{\mathcal{I}_{\check{\mathbf{C}}}} \mathbf{P}_{\check{\mathcal{L}}}(\check{\mathbf{N}}_{\mathbf{C}})$, and $\check{\mathbf{N}}^+ = \check{\mathbf{N}} - \mathbf{P}_{\mathcal{I}_{\check{\mathbf{C}}}} \mathbf{P}_{\check{\mathcal{L}}}(\check{\mathbf{N}})$. It is shown in [9] (Lemma 11) that for any $\mathbf{X} \in \mathbb{R}^{m \times \check{n}_2}$

$$\mathbf{P}_{\mathcal{I}_{\check{\mathbf{C}}}} \mathbf{P}_{\overline{\mathbf{V}}} \mathbf{P}_{\mathcal{I}_{\check{\mathbf{C}}}}(\mathbf{X}) = \mathbf{X}(\mathbf{P}_{\mathcal{I}_{\check{\mathbf{C}}}}(\overline{\mathbf{V}}^T))^T \mathbf{P}_{\mathcal{I}_{\check{\mathbf{C}}}}(\overline{\mathbf{V}}^T),$$

and correspondingly for some $\psi$, we have

$$\|\mathbf{P}_{\mathcal{I}_{\check{\mathbf{C}}}} \mathbf{P}_{\overline{\mathbf{V}}} \mathbf{P}_{\mathcal{I}_{\check{\mathbf{C}}}}(\mathbf{X})\|_2 = \|\mathbf{X}(\mathbf{P}_{\mathcal{I}_{\check{\mathbf{C}}}}(\overline{\mathbf{V}}^T))^T \mathbf{P}_{\mathcal{I}_{\check{\mathbf{C}}}}(\overline{\mathbf{V}}^T)\|_2$$
$$\leq \|\mathbf{X}\|_2 \|(\mathbf{P}_{\mathcal{I}_{\check{\mathbf{C}}}}(\overline{\mathbf{V}}^T))^T \mathbf{P}_{\mathcal{I}_{\check{\mathbf{C}}}}(\overline{\mathbf{V}}^T)\|_2 \leq \psi \|\mathbf{X}\|_2, \quad (58)$$

where the last inequality follows from the bound $\|(\mathbf{P}_{\mathcal{I}_{\check{\mathbf{C}}}}(\overline{\mathbf{V}}^T))^T \mathbf{P}_{\mathcal{I}_{\check{\mathbf{C}}}}(\overline{\mathbf{V}}^T)\|_2 \leq \psi$. It is shown in [9] that if $\check{k} = |\mathcal{I}_{\check{\mathbf{C}}}|$ satisfies (26), then we have $\psi < \frac{1}{4}$.

The main body of the proof is to construct a dual certificate to guarantee that the optimal solution pair of (29) is "close" to a pair of $(\check{\mathbf{L}}_0, \check{\mathbf{C}}_0)$, which has the correct column space and the correct column support respectively, in terms of the spectral norm, given the $\ell_{1,2}$-norm of the noise term. We demonstrate this in Lemma VI.8.

**Lemma VI.8** (Adapted from Theorem 5 in [9]). *Let $\widehat{\mathbf{L}}$, $\widehat{\mathbf{C}}$ be an optimal solution pair of (29). Suppose $\lambda = \frac{\sqrt{9+1024\mu_{\check{\mathbf{L}}}r}}{14\sqrt{\tilde{n}_2}} < 1$ and $\psi < \frac{1}{4}$. Let $\check{\mathbf{M}}_0 = \check{\mathbf{L}}_1 + \check{\mathbf{C}}_1$, where $\mathbf{P}_{\check{\mathcal{L}}}(\check{\mathbf{L}}_1) = \check{\mathbf{L}}_1$ and $\mathbf{P}_{\mathcal{I}_{\check{\mathbf{C}}}}(\check{\mathbf{C}}_1) = \check{\mathbf{C}}_1$. If there exists $\mathbf{Q}$ such that*

$$\mathbf{P}_{\mathcal{T}(\check{\mathbf{L}}_1)}(\mathbf{Q}) = \Re(\check{\mathbf{L}}_1), \quad \|\mathbf{P}_{\mathcal{T}(\check{\mathbf{L}}_1)^\perp}(\mathbf{Q})\|_2 \leq 1/2,$$
$$\mathbf{P}_{\mathcal{I}_{\check{\mathbf{C}}}}(\mathbf{Q})/\lambda \in \mathfrak{G}(\check{\mathbf{C}}_1), \quad \|\mathbf{P}_{\mathcal{I}_{\check{\mathbf{C}}}^c}(\mathbf{Q})\|_{\infty,2} \leq \lambda/2,$$

*then these exists a pair $(\check{\mathbf{L}}_0, \check{\mathbf{C}}_0)$ such that $\check{\mathbf{M}}_0 = \check{\mathbf{L}}_0 + \check{\mathbf{C}}_0$, $\mathbf{P}_{\check{\mathcal{L}}}(\check{\mathbf{L}}_0) = \check{\mathbf{L}}_0$ and $\mathbf{P}_{\mathcal{I}_{\check{\mathbf{C}}}}(\check{\mathbf{C}}_0) = \check{\mathbf{C}}_0$, and*

$$\|\widehat{\mathbf{L}} - \check{\mathbf{L}}_0\|_2 \leq 10\|\check{\mathbf{N}}\|_{1,2}, \quad \|\widehat{\mathbf{C}} - \check{\mathbf{C}}_0\|_2 \leq 9\|\check{\mathbf{N}}\|_{1,2}.$$

*Proof of Lemma VI.8.* Let us introduce two quantities $\mathbf{W}$ and $\mathbf{F}$ that are related to the subgradient of $\|\widehat{\mathbf{L}}_1\|_*$ and $\|\widehat{\mathbf{C}}_1\|_{1,2}$. We have from Theorem 3 in [9] that for any fixed perturbation $\mathbf{\Delta} \neq \mathbf{0}$, $(\widehat{\mathbf{L}}_1 + \mathbf{\Delta}, \widehat{\mathbf{C}}_1 - \mathbf{\Delta})$ is strictly worse than $(\widehat{\mathbf{L}}_1, \widehat{\mathbf{C}}_1)$, unless $\mathbf{\Delta} \in \mathbf{P}_{\check{\mathcal{L}}} \cap \mathbf{P}_{\mathcal{I}_{\check{\mathbf{C}}}}$. Let $\mathbf{W}$ be such that $\|\mathbf{W}\|_2 = 1$, $\langle \mathbf{W}, \mathbf{P}_{\mathcal{T}(\widehat{\mathbf{L}})^\perp}(\mathbf{\Delta}) \rangle = \|\mathbf{P}_{\mathcal{T}(\widehat{\mathbf{L}})^\perp}(\mathbf{\Delta})\|_*$, and $\mathbf{P}_{\mathcal{T}(\widehat{\mathbf{L}})}(\mathbf{W}) = \mathbf{0}$. Let $\mathbf{F}$ be such that

$$\mathbf{F}_{:,i} = \begin{cases} \frac{-\mathbf{\Delta}_{:,i}}{\|\mathbf{\Delta}_{:,i}\|_2}, & \text{if } i \notin \mathcal{I}_\mathbf{C}, \text{ and } \mathbf{\Delta}_{:,i} \neq \mathbf{0} \\ \mathbf{0}, & \text{otherwise.} \end{cases}$$

Then $\mathbf{P}_{\mathcal{T}(\widehat{\mathbf{L}})}(\mathbf{Q}) + \mathbf{W}$ is a subgradient of $\|\widehat{\mathbf{L}}_1\|_*$ and $\mathbf{P}_{\mathcal{I}_\mathbf{C}}(\mathbf{Q})/\lambda + \mathbf{F}$ is a subgradient of $\|\widehat{\mathbf{C}}_1\|_{1,2}$.

From the optimality of $\widehat{\mathbf{L}}$ and $\widehat{\mathbf{C}}$, we have

$$\|\check{\mathbf{L}}_1\|_* + \lambda\|\check{\mathbf{C}}_1\|_{1,2} \geq \|\widehat{\mathbf{L}}_1\|_* + \lambda\|\widehat{\mathbf{C}}_1\|_{1,2}$$
$$\geq \|\check{\mathbf{L}}_1\|_* + \lambda\|\check{\mathbf{C}}_1\|_{1,2} + \langle \mathbf{P}_{\mathcal{T}(\check{\mathbf{L}}_1)}(\mathbf{Q}) + \mathbf{W}, \check{\mathbf{N}}_\mathbf{L} \rangle$$
$$+ \lambda\langle \mathbf{P}_{\mathcal{I}_\mathbf{C}}(\mathbf{Q})/\lambda + \mathbf{F}, \check{\mathbf{N}}_\mathbf{C} \rangle$$
$$\stackrel{(i)}{=} \|\check{\mathbf{L}}_1\|_* + \lambda\|\check{\mathbf{C}}_1\|_{1,2} + \|\mathbf{P}_{\mathcal{T}(\widehat{\mathbf{L}})^\perp}(\check{\mathbf{N}}_\mathbf{L})\|_* + \lambda\|\mathbf{P}_{\mathcal{I}_{\check{\mathbf{C}}}^c}(\check{\mathbf{N}}_\mathbf{C})\|_{1,2}$$
$$+ \langle \mathbf{P}_{\mathcal{T}(\check{\mathbf{L}}_1)}(\mathbf{Q}), \check{\mathbf{N}}_\mathbf{L} \rangle + \langle \mathbf{P}_{\mathcal{I}_\mathbf{C}}(\mathbf{Q}), \check{\mathbf{N}}_\mathbf{C} \rangle$$
$$= \|\check{\mathbf{L}}_1\|_* + \lambda\|\check{\mathbf{C}}_1\|_{1,2} + \|\mathbf{P}_{\mathcal{T}(\widehat{\mathbf{L}})^\perp}(\check{\mathbf{N}}_\mathbf{L})\|_* + \lambda\|\mathbf{P}_{\mathcal{I}_{\check{\mathbf{C}}}^c}(\check{\mathbf{N}}_\mathbf{C})\|_{1,2}$$
$$- \langle \mathbf{P}_{\mathcal{T}(\check{\mathbf{L}}_1)^\perp}(\mathbf{Q}), \check{\mathbf{N}}_\mathbf{L} \rangle - \langle \mathbf{P}_{\mathcal{I}_{\check{\mathbf{C}}}^c}(\mathbf{Q}), \check{\mathbf{N}}_\mathbf{C} \rangle + \langle \mathbf{Q}, \check{\mathbf{N}}_\mathbf{L} + \check{\mathbf{N}}_\mathbf{C} \rangle$$
$$\geq \|\check{\mathbf{L}}_1\|_* + \lambda\|\check{\mathbf{C}}_1\|_{1,2} + (1 - \|\mathbf{P}_{\mathcal{T}(\check{\mathbf{L}}_1)^\perp}(\mathbf{Q})\|)\|\mathbf{P}_{\mathcal{T}(\check{\mathbf{L}}_1)^\perp}(\check{\mathbf{N}}_\mathbf{L})\|_*$$
$$+ (\lambda - \|\mathbf{P}_{\mathcal{I}_{\check{\mathbf{C}}}^c}(\mathbf{Q})\|_{\infty,2})\|\mathbf{P}_{\mathcal{I}_{\check{\mathbf{C}}}^c}(\check{\mathbf{N}}_\mathbf{C})\|_{1,2} + \langle \mathbf{Q}, \mathbf{N} \rangle$$
$$\geq \|\check{\mathbf{L}}_1\|_* + \lambda\|\check{\mathbf{C}}_1\|_{1,2} + \frac{1}{2}\|\mathbf{P}_{\mathcal{T}(\check{\mathbf{L}}_1)^\perp}(\check{\mathbf{N}}_\mathbf{L})\|_*$$
$$+ \frac{\lambda}{2}\|\mathbf{P}_{\mathcal{I}_{\check{\mathbf{C}}}^c}(\check{\mathbf{N}}_\mathbf{C})\|_{1,2} - \|\check{\mathbf{N}}\|_{1,2}\|\mathbf{Q}\|_{\infty,2}, \quad (59)$$

where $(i)$ is from the choice of $\mathbf{W}$ and $\mathbf{F}$ above. From (59), we have

$$\|\mathbf{P}_{\mathcal{T}(\check{\mathbf{L}}_1)^\perp}(\check{\mathbf{N}}_\mathbf{L})\|_2 \leq \|\mathbf{P}_{\mathcal{T}(\check{\mathbf{L}}_1)^\perp}(\check{\mathbf{N}}_\mathbf{L})\|_*$$
$$\leq 2\lambda\|\check{\mathbf{N}}\|_{1,2}\|\mathbf{Q}\|_{\infty,2} \leq 2\lambda\|\check{\mathbf{N}}\|_{1,2}, \quad (60)$$
$$\|\mathbf{P}_{\mathcal{I}_{\check{\mathbf{C}}}^c}(\check{\mathbf{N}}_\mathbf{C})\|_2 \leq \|\mathbf{P}_{\mathcal{I}_{\check{\mathbf{C}}}^c}(\check{\mathbf{N}}_\mathbf{C})\|_{1,2}$$
$$\leq 2\|\check{\mathbf{N}}\|_{1,2}\|\mathbf{Q}\|_{\infty,2} \leq 2\|\check{\mathbf{N}}\|_{1,2}. \quad (61)$$

From the result in [9] (eqn. (16)), we have

$$\mathbf{P}_{\mathcal{I}_{\check{\mathbf{C}}}}(\check{\mathbf{N}}_\mathbf{C}^+) = \mathbf{P}_{\mathcal{I}_{\check{\mathbf{C}}}}(\check{\mathbf{N}}) - \mathbf{P}_{\mathcal{I}_{\check{\mathbf{C}}}}\mathbf{P}_{\mathcal{T}(\check{\mathbf{L}}_1)^\perp}(\check{\mathbf{N}}_\mathbf{L}) - \mathbf{P}_{\mathcal{I}_{\check{\mathbf{C}}}}\mathbf{P}_{\mathcal{T}(\check{\mathbf{L}}_1)}(\check{\mathbf{N}})$$
$$+ \mathbf{P}_{\mathcal{I}_{\check{\mathbf{C}}}}\mathbf{P}_{\mathcal{T}(\check{\mathbf{L}}_1)}\mathbf{P}_{\mathcal{I}_{\check{\mathbf{C}}}^c}(\check{\mathbf{N}}_\mathbf{C}) + \mathbf{P}_{\mathcal{I}_{\check{\mathbf{C}}}}\mathbf{P}_{\overline{\mathbf{V}}}\mathbf{P}_{\mathcal{I}_{\check{\mathbf{C}}}}(\check{\mathbf{N}}_\mathbf{C}^+).$$

By triangle inequality, the equality above implies

$$\|\mathbf{P}_{\mathcal{I}_{\check{\mathbf{C}}}}(\check{\mathbf{N}}_\mathbf{C}^+)\|_2$$
$$\leq \|\mathbf{P}_{\mathcal{I}_{\check{\mathbf{C}}}}(\check{\mathbf{N}}) - \mathbf{P}_{\mathcal{I}_{\check{\mathbf{C}}}}\mathbf{P}_{\mathcal{T}(\check{\mathbf{L}}_1)}(\check{\mathbf{N}})\|_2 + \|\mathbf{P}_{\mathcal{I}_{\check{\mathbf{C}}}}\mathbf{P}_{\mathcal{T}(\check{\mathbf{L}}_1)^\perp}(\check{\mathbf{N}}_\mathbf{L})\|_2$$
$$+ \|\mathbf{P}_{\mathcal{I}_{\check{\mathbf{C}}}}\mathbf{P}_{\mathcal{T}(\check{\mathbf{L}}_1)}\mathbf{P}_{\mathcal{I}_{\check{\mathbf{C}}}^c}(\check{\mathbf{N}}_\mathbf{C})\|_2 + \|\mathbf{P}_{\mathcal{I}_{\check{\mathbf{C}}}}\mathbf{P}_{\overline{\mathbf{V}}}\mathbf{P}_{\mathcal{I}_{\check{\mathbf{C}}}}(\check{\mathbf{N}}_\mathbf{C}^+)\|_2$$
$$\stackrel{(i)}{\leq} \|\check{\mathbf{N}}\|_2 + \|\mathbf{P}_{\mathcal{T}(\check{\mathbf{L}}_1)^\perp}(\check{\mathbf{N}}_\mathbf{L})\|_2 + \|\mathbf{P}_{\mathcal{I}_{\check{\mathbf{C}}}^c}(\check{\mathbf{N}}_\mathbf{C})\|_2$$
$$+ \psi\|\mathbf{P}_{\mathcal{I}_{\check{\mathbf{C}}}}(\check{\mathbf{N}}_\mathbf{C}^+)\|_2$$
$$\stackrel{(ii)}{\leq} (1 + 2\lambda + 2)\|\check{\mathbf{N}}\|_{1,2} + \psi\|\mathbf{P}_{\mathcal{I}_{\check{\mathbf{C}}}}(\check{\mathbf{N}}_\mathbf{C}^+)\|_2, \quad (62)$$

where $(i)$ is from (58), and $(ii)$ is from (60), (61), and the fact $\|\check{\mathbf{N}}\|_2 \leq \|\check{\mathbf{N}}\|_F \leq \|\check{\mathbf{N}}\|_{1,2}$. From (62), we have

$$\|\mathbf{P}_{\mathcal{I}_{\check{\mathbf{C}}}}(\check{\mathbf{N}}_\mathbf{C}^+)\|_2 \leq \frac{(1 + 2\lambda\sqrt{\tilde{n}_2} + 2\sqrt{\tilde{n}_2})\|\check{\mathbf{N}}\|_{1,2}}{1 - \psi}. \quad (63)$$

Combining (61), (63), and the fact that $\lambda < 1$ and $\psi < \frac{1}{4}$, we have

$$\|\check{\mathbf{N}}_\mathbf{C}^+\|_2 = \|\mathbf{P}_{\mathcal{I}_{\check{\mathbf{C}}}^c}(\check{\mathbf{N}}_\mathbf{C}) + \mathbf{P}_{\mathcal{I}_{\check{\mathbf{C}}}}(\check{\mathbf{N}}_\mathbf{C}^+)\|_{1,2}$$
$$\leq \|\mathbf{P}_{\mathcal{I}_{\check{\mathbf{C}}}^c}(\check{\mathbf{N}}_\mathbf{C})\|_2 + \|\mathbf{P}_{\mathcal{I}_{\check{\mathbf{C}}}}(\check{\mathbf{N}}_\mathbf{C}^+)\|_2 \leq 9\|\check{\mathbf{N}}\|_{1,2}.$$

Finally, note that we have

$$\check{\mathbf{N}}_\mathbf{C}^+ = (\mathbf{I} - \mathbf{P}_{\mathcal{I}_{\check{\mathbf{C}}}}\mathbf{P}_{\mathcal{I}_{\check{\mathcal{L}}}})(\widehat{\mathbf{C}} - \check{\mathbf{C}}_1) = \widehat{\mathbf{C}} - \check{\mathbf{C}}_0,$$

where $\check{\mathbf{C}}_0 = \check{\mathbf{C}}_1 + \mathbf{P}_{\mathcal{I}_{\check{\mathbf{C}}}}\mathbf{P}_{\mathcal{I}_{\check{\mathcal{L}}}}(\widehat{\mathbf{C}} - \check{\mathbf{C}}_1)$. Since $\check{\mathbf{C}}_1 \in \mathcal{I}_{\check{\mathbf{C}}}$, this implies $\check{\mathbf{C}}_0 \in \mathcal{I}_{\check{\mathbf{C}}}$ and

$$\|\widehat{\mathbf{C}} - \check{\mathbf{C}}_0\|_2 \leq 9\|\check{\mathbf{N}}\|_{1,2}.$$

Further let $\check{\mathbf{L}}_0 = \check{\mathbf{L}}_1 - \mathbf{P}_{\mathcal{I}_{\check{\mathbf{C}}}}\mathbf{P}_{\mathcal{I}_{\check{\mathcal{L}}}}(\widehat{\mathbf{C}} - \check{\mathbf{C}}_1)$, we have that $\check{\mathbf{L}}_0$ and $\check{\mathbf{C}}_0$ are a pair of successful decomposition, and

$$\|\check{\mathbf{L}}_0 - \widehat{\mathbf{L}}\|_2 \leq \|\check{\mathbf{N}}\|_{1,2} + \|\widehat{\mathbf{C}} - \check{\mathbf{C}}_0\|_2 \leq 10\|\check{\mathbf{N}}\|_{1,2}.$$

□

### H. Further Intermediate Results

In this section, we provide the statement of several intermediate results adopted in our analysis.

The first intermediate result provides the additive perturbation bound for the singular values.

**Lemma VI.9** (Theorem 1 of [44]). *Suppose $n_1 \leq n_2$. Let $\mathbf{A} \in \mathbb{R}^{n_1 \times n_2}$ be a matrix with singular values $\{\sigma_i\}_{i=1}^{n_1}$, and $\widetilde{\mathbf{A}} = \mathbf{A} + \mathbf{E}$ be a perturbation of $\mathbf{A}$ with singular values $\{\widetilde{\sigma}_i\}_{i=1}^{n_1}$. Then for any $i \in [n_1]$, the following bound holds:*

$$|\widetilde{\sigma}_i - \sigma_i| \leq \|\mathbf{E}\|_2.$$

The second intermediate result provides the additive perturbation bound for the orthogonal projection.

**Lemma VI.10** (Adapted from Theorem 2.2 of [45]). *Let $\mathbf{A} \in \mathbb{R}^{n_1 \times n_2}$ be a rank-$r$ matrix with SVD*

$$\mathbf{A} = [\mathbf{U}_1 \mathbf{U}_2] \begin{bmatrix} \mathbf{\Sigma}_1 & \mathbf{0} \\ \mathbf{0} & \mathbf{0} \end{bmatrix} \begin{bmatrix} \mathbf{V}_1^T \\ \mathbf{V}_2^T \end{bmatrix},$$

*and $\widetilde{\mathbf{A}} = \mathbf{A} + \mathbf{E}$ be a perturbation of $\mathbf{A}$ with $\mathrm{rank}(\widetilde{\mathbf{A}}) = \mathrm{rank}(\mathbf{A}) = r$. Let $\widetilde{\mathcal{A}}$ and $\mathcal{A}$ be the column spaces of $\widetilde{\mathbf{A}}$ and $\mathbf{A}$ respectively. Then the following bound holds:*

$$\|\mathbf{P}_{\widetilde{\mathcal{A}}} - \mathbf{P}_{\mathcal{A}}\|_2 \leq \min\{1, \|\mathbf{A}^\dagger\|_2\|\mathbf{E}\mathbf{V}_1\|_2, \|\widetilde{\mathbf{A}}^\dagger\|_2\|\mathbf{U}_2^T\mathbf{E}\|_2\},$$

*where $\mathbf{A}^\dagger$ is the Moore-Penrose inverse of $\mathbf{A}$.*

## REFERENCES

[1] B. Mehta and W. Nejdl, "Attack resistant collaborative filtering," in *Proceedings of the 31st annual international ACM SIGIR conference on Research and development in information retrieval*. ACM, 2008, pp. 75–82.

[2] A. Lakhina, M. Crovella, and C. Diot, "Diagnosing network-wide traffic anomalies," in *ACM SIGCOMM Computer Communication Review*. ACM, 2004, vol. 34, pp. 219–230.

[3] L. Itti, C. Koch, and E. Niebur, "A model of saliency-based visual attention for rapid scene analysis," *IEEE Trans. Pattern Analysis and Machine Intelligence*, vol. 20, no. 11, pp. 1254–1259, 1998.

[4] J. Harel, C. Koch, and P. Perona, "Graph-based visual saliency.," in *Proc. NIPS*, 2006, pp. 545–552.

[5] T. Liu, J. Sun, N. Zheng, X. Tang, and H. Shum, "Learning to detect a salient object," in *Proc. CVPR*, 2007.

[6] X. Shen and Y. Wu, "A unified approach to salient object detection via low rank matrix recovery," in *Proc. CVPR*, 2012, pp. 853–860.

[7] V. Chandrasekaran, S. Sanghavi, P. Parrilo, and A. Willsky, "Rank-sparsity incoherence for matrix decomposition," *SIAM J. Optimization*, vol. 21, no. 2, pp. 572–596, 2011.

[8] E. J. Candès, X. Li, Y. Ma, and J. Wright, "Robust principal component analysis?," *J. ACM*, vol. 58, no. 3, pp. 11:1–11:37, 2011.

[9] H. Xu, C. Caramanis, and S. Sanghavi, "Robust PCA via outlier pursuit," *IEEE Trans. Inform. Theory*, vol. 58, no. 5, pp. 3047–3064, 2012.

[10] M. McCoy and J. A. Tropp, "Two proposals for robust PCA using semidefinite programming," *Electronic J. of Statistics*, vol. 5, pp. 1123–1160, 2011.

[11] M. Soltanolkotabi and E. Candes, "A geometric analysis of subspace clustering with outliers," *The Annals of Statistics*, vol. 40, no. 4, pp. 2195–2238, 2012.

[12] M. Hardt and A. Moitra, "Algorithms and hardness for robust subspace recovery," in *Conf. on Learning Theory*, 2013, pp. 354–375.

[13] G. Lerman, M. B. McCoy, J. A. Tropp, and T. Zhang, "Robust computation of linear models by convex relaxation," *Foundations of Computational Mathematics*, pp. 1–48, 2014.

[14] Y. Chen, H. Xu, C. Caramanis, and S. Sanghavi, "Matrix completion with column manipulation: Near-optimal sample-robustness-rank tradeoffs," *IEEE Trans. Inform. Theory*, vol. 62, no. 1, pp. 503–526, 2016.

[15] X. Li and J. Haupt, "Identifying outliers in large matrices via randomized adaptive compressive sampling," *Trans. Signal Processing*, vol. 63, no. 7, pp. 1792–1807, 2015.

[16] X. Li and J. Haupt, "A refined analysis for the sample complexity of adaptive compressive outlier sensing," in *IEEE Workshop on Statistical Signal Processing*, 2016.

[17] E. J. Candès, J. Romberg, and T. Tao, "Robust uncertainty principles: Exact signal recovery from highly incomplete frequency information," *IEEE Trans. Inform. Theory*, vol. 52, no. 2, pp. 489–509, 2006.

[18] D. Donoho, "Compressed sensing," *IEEE Trans. Inform. Theory*, vol. 52, no. 4, pp. 1289–1306, 2006.

[19] E. Bashan, R. Raich, and A. O. Hero, "Optimal two-stage search for sparse targets using convex criteria," *IEEE Trans Signal Processing*, vol. 56, no. 11, pp. 5389–5402, 2008.

[20] J. Haupt, R. M. Castro, and R. Nowak, "Distilled sensing: Adaptive sampling for sparse detection and estimation," *IEEE Trans. Information Theory*, vol. 57, no. 9, pp. 6222–6235, 2011.

[21] E. Bashan, G. Newstadt, and A. O. Hero, "Two-stage multiscale search for sparse targets," *IEEE Trans Signal Processing*, vol. 59, no. 5, pp. 2331–2341, 2011.

[22] M. L. Malloy and R. Nowak, "Sequential testing for sparse recovery," *arXiv preprint:1212.1801*, 2012.

[23] R. M. Castro, "Adaptive sensing performance lower bounds for sparse signal detection and support estimation," *arXiv preprint:1206.0648*, 2012.

[24] A. Krishnamurthy, J. Sharpnack, and A. Singh, "Recovering graph-structured activations using adaptive compressive measurements," *arXiv preprint:1305.0213*, 2013.

[25] D. Wei and A. O. Hero, "Multistage adaptive estimation of sparse signals," *IEEE J. of Selected Topics in Signal Processing*, vol. 7, no. 5, pp. 783–796, 2013.

[26] Y. Chen, H. Xu, C. Caramanis, and S. Sanghavi, "Robust matrix completion with corrupted columns," *arXiv preprint:1102.2254*, 2011.

[27] P. Huber, *Robust statistics*, Springer, 2011.

[28] H. Nyquist, "Least orthogonal absolute deviations," *Computational Statistics & Data Analysis*, vol. 6, no. 4, pp. 361–367, 1988.

[29] C. Yang, D. Robinson, and R. Vidal, "Sparse subspace clustering with missing entries," in *Proceedings of The 32nd International Conference on Machine Learning*, 2015, pp. 2463–2472.

[30] M. Rahmani and G. Atia, "Randomized robust subspace recovery for high dimensional data matrices," *arXiv preprint arXiv:1505.05901*, 2015.

[31] Y. She, S. Li, and D. Wu, "Robust orthogonal complement principal component analysis," *arXiv preprint arXiv:1410.1173*, 2014.

[32] T. Zhang and G. Lerman, "A novel m-estimator for robust pca," *J. Machine Learning Research*, vol. 15, no. 1, pp. 749–808, 2014.

[33] O. Klopp, K. Lounici, and A. Tsybakov, "Robust matrix completion," *arXiv preprint arXiv:1412.8132*, 2014.

[34] G. Obozinski, M. J Wainwright, and M. Jordan, "Support union recovery in high-dimensional multivariate regression," *The Annals of Statistics*, pp. 1–47, 2011.

[35] X. Li and J. Haupt, "Locating salient group-structured image features via adaptive compressive sensing," in *GlobalSIP*, 2015.

[36] X. Li and J. Haupt, "Outlier identification via randomized adaptive compressive sampling," in *ICASSP*, 2015.

[37] D. Woodruff, "Sketching as a tool for numerical linear algebra," *Found. Trends Theor. Comput. Sci.*, vol. 10, no. 1–2, pp. 1–157, Oct. 2014.

[38] A. C. Gilbert, J. Y. Park, and M. B. Wakin, "Sketched SVD: Recovering spectral features from compressive measurements," *arXiv preprint:1211.0361*, 2012.

[39] A. Klenke and L. Mattner, "Stochastic ordering of classical discrete distributions," *Advances in Applied probability*, vol. 42, no. 2, pp. 392–410, 2010.

[40] I. Johnstone, "Chi-square oracle inequalities," *Lecture Notes-Monograph Series*, pp. 399–418, 2001.

[41] C. McDiarmid, "Concentration," in *Probabilistic methods for algorithmic discrete mathematics*, pp. 195–248. Springer, 1998.

[42] A. Krishnamurthy and A. Singh, "On the power of adaptivity in matrix completion and approximation," *arXiv preprint:1407.3619*, 2014.

[43] J. A. Tropp, "User-friendly tail bounds for sums of random matrices," *Foundations of Computational Mathematics*, vol. 12, no. 4, pp. 389–434, 2012.

[44] G. Stewart, "Perturbation theory for the singular value decomposition," 1998.

[45] B. Li, W. Li, and L. Cui, "New bounds for perturbation of the orthogonal projection," *Calcolo*, vol. 50, no. 2, pp. 69—78, 2013.